\magnification\magstep1
\overfullrule=0pt

{\catcode`p =12 \catcode`t =12 \gdef\eeaa#1pt{#1}}
\def\accentadj#1{\setbox0\hbox{$#1$}\kern
		\expandafter\eeaa\the\fontdimen1\textfont1 \ht0 }
\def\overchar#1#2{\vbox{\ialign{$\hfil##\hfil$\crcr	%
	\accentadj{#2}#1\crcr\noalign{\kern.37ex\nointerlineskip}
	\displaystyle#2\crcr}}}
\def\dotskip{\mskip-1.6\thinmuskip}
%
%
\def\DDD#1{\overchar{\ldotp\dotskip\ldotp
	\dotskip\ldotp}{#1}}

\def\dddot#1{\overchar{\ldotp\dotskip\ldotp
	\dotskip\ldotp}{#1}}

\def\SU{\mathcode`,="8000 \mathcode`!="8000\relax}	
{\catcode`\,=\active \catcode`\!=\active
\gdef,{\mkern-2mu{}_} \gdef!{\mkern-2mu{}^}}
\def\Ko{\oalign{$K$\cr	
	\noalign{\vskip0.3ex}	
	\hidewidth$\scriptstyle0$\hidewidth}{}}
\def\Ro{\oalign{$R$\cr
\noalign{\vskip0.3ex}
\hidewidth$\scriptstyle0$\hidewidth}{}}

\def\DD{\ddot}

\def\D {\dot }

\def\1{\oalign{$\hat{{\it L\,\,}}$\cr
\noalign{\vskip0.3ex}
\hidewidth$\xi_1$\hidewidth}{}}

\def\2{\oalign{$\hat{{\it L\,\,}}$\cr
\noalign{\vskip0.3ex}
\hidewidth$\xi_2$\hidewidth}{}}

 \def\Po{\oalign{$P$\cr
\noalign{\vskip0.3ex}
\hidewidth$\scriptstyle0$\hidewidth}{}}

\baselineskip 16pt

\def\R{\SU R^a,{bcd} }

\def\Po{$\SU \Po^a,{bcd}$ }

\def\Ko{$\SU \Ko^a,{bcd}$ }

\def\Ro{$\SU \Ro^a,{bcd}$ }

\def\R*{\check R}

\def\B*{\check B}

\def\Q*{\check Q}

\parindent=0pt

\def\PPh#1{\setbox0\hbox{$#1\rm I$}\mathord{\vcenter{\ialign{$#1\rm##$\cr
I\cr\noalign{\nointerlineskip \vskip-0.541\ht0}P\cr}}}}

\def\Ph{{\mathpalette\PPh{}}}
\def\Ed{{\mathord{\mkern5mu\mathaccent"7020{\mkern-5mu\partial}}}}

\font\mybold=cmmib10
\chardef\Myxi="18
\def\boldxi{\hbox{\mybold\Myxi}}

\parindent=0pt

\

\centerline{}
\centerline{\bf    INTEGRATION  IN THE GHP FORMALISM II: AN OPERATOR APPROACH
 }\smallskip
\centerline{\bf FOR SPACETIMES WITH KILLING VECTORS, WITH APPLICATIONS}
\smallskip \centerline{\bf TO
TWISTING TYPE N SPACES.}

\

\

\centerline{S. Brian Edgar\footnote{$^1$}{ Department of Mathematics,
 Link\"oping University, Link\"oping, Sweden S-581 83.
 E-mail: bredg@math.liu.se. FAX: +46 13 100746.} and Garry
Ludwig\footnote{$^2$}
{ Department of Mathematical Sciences, University of Alberta,
Edmonton, Alberta,
Canada  T6G 2G1. E-mail: Garry.Ludwig@UAlberta.CA} }

\bigskip

\

\

\vfill\eject

\beginsection ABSTRACT.

Held has proposed a coordinate- and gauge-free integration procedure within the
GHP formalism built around four functionally independent zero-weighted scalars
constructed from the spin coefficients and the Riemann tensor
components. Unfortunately, a spacetime with Killing vectors (and hence cyclic
coordinates in the metric, and in all quantities constructed from the metric)
will be unable to supply the full quota of four scalars of this type.  However,
for such a spacetime additional scalars are supplied by the components of the
Killing vectors; by using these alongside the spin coefficients and the Riemann
tensor components we have the possibility of constructing the full quota of four
functionally independent zero-weighted scalars, and of exploiting Held's
procedure.

As an illustration we investigate the vacuum Type N spaces admitting a Killing
vector and a homothetic Killing vector. We show how the properties of
separability, redundancy, decoupling and reduction of order can be understood
and interpreted in a very general manner in our approach, and the advantages of
being able to postpone the explicit coordinate choice to the very last step,
when it can be used to simplify the residual ordinary differential equations. In
a direct manner, we reduce the problem to a pair of ordinary differential
operator `master equations', making use of a new zero-weighted GHP operator. By
first rewriting the master equations as a closed set of complex first order
equations, we reduce the problem to one real third order operator differential
equation for a complex function of a real variable --- but with still the
freedom to choose explicitly our fourth coordinate.
 It is then easy to see there are a whole class of coordinate choices where the
problem reduces essentially to one real third order differential equation for a
real function of a real variable. An alternative, more algorithmic approach,
using a closed chain of real first order equations for real functions, reduces
the problem to the same order, but in a more natural and much more concise
form. It is also outlined how the various other third order differential
equations, which have been derived previously in work on this problem, can be
deduced from our master equations.

\

\vfill\eject

\beginsection 1. Introduction.

In a previous paper [1] we developed and illustrated a coordinate-dependent
integration procedure in the GHP formalism [2], but, in spite of its
advantages, especially in efficiency, over the analogous NU [3,4] approach in
the
NP formalism [5], we feel that it does not exploit the GHP formalism to its
full potential. We also
noted there an alternative coordinate-independent, operator approach advocated
and developed by Held [6-10], and modified and illustrated recently in [11]. In
this present paper we will
develop this approach further --- specifically  for spacetimes admitting at
least one Killing vector --- and illustrate it by an application    to the same
problem as considered in the previous paper, [1] --- the $G_2 $ class of the
twisting type N vacuum spaces (NT spaces).  We emphasise however,  that both
papers are self-contained.

In the remainder of this section we will outline this GHP operator-integration
procedure --- first of all
in general, and
then specialised  to the particular problem to be considered.

\

\beginsection 1.1. Outline of the  integration procedure: the general case.

As discussed in detail in the introduction in [1], the GHP formalism
consists of a complete system
of  three sets of equations: the GHP commutator equations, the
GHP Ricci equations, and the GHP Bianchi equations. A procedure for
integrating this system was  summarised  in five steps in [11]:

The first two steps are a coordinate-free integration procedure for the
operator $\Ph $, exactly as proposed by Held, [6]; this $\rho $-integration
process is a generalisation of the $r $-integration process in the NP
formalism.
The third step involves the application of the commutator equations to  three
complex quantities --- two  zero-weighted complex quantities which supply four
functionally-independent zero-weighted real quantities and one weighted (by
which we shall mean neither weight being zero,  i.e. $s\ne 0\ne t $ or
equivalently $p\ne \pm q $, in the usual notations, [2]) complex quantity ---
so
that the commutator equations are replaced by an alternative set of equations;
we shall refer to these four real functionally-independent zero-weighted
quantities as $`$coordinate candidates' because, at the last two steps, they
will  usually be the obvious (but not necessarily always the most suitable)
choice for the four coordinates, whereas the one weighted complex quantity will
usually be transformed to unity by choice of gauge. We can refer to this new
set
of equations as the `GHP metric candidate equations', but  emphasise that
although  this new set of equations is constructed in a manner analogous  to
the
NP metric equations in the NP-NU procedure, and to the GHP metric equations in
the earlier paper, [1] there is a fundamental difference; these coordinate
candidates
 are chosen from within the system of equations, motivated only by
 considerations of their mathematical structure, whereas in [1]
choices were imposed from the outside, motivated largely by physical and
external geometric considerations. Another important difference is that many
equations in this new set turn out to be identities, modulo the  equations in
the other sets; in fact we deliberately choose our coordinate candidates with
the intention of  achieving  such structural simplifications.

After these three steps we will have
reduced the complete system  to a much smaller
 sufficient subsystem of differential (operator) equations; this subsystem will
consist of (the equivalent of) six real tables for the action of the
four GHP operators on each of the four (real) coordinate candidates and on the
one complex weighted quantity, together with any residual differential
equations
from the original system of equations.

The last two steps involve the
introduction of an explicit coordinate system and the adoption  of a specific
tetrad gauge; these
choices  will be made in such a manner that the  reduced sufficient subsystem
of
differential equations become as simple and manageable as possible. If we
choose
our four coordinate candidates to be our coordinates, then the four real tables
for these quantities simply become the definitions of the four differential
operators (equivalently, the tetrad vectors)  in this choice of coordinate
system; if we choose to use our gauge freedom to reduce the  weighted complex
quantity to unity, then the complex
 table for this  quantity simply becomes the definitions of the badly
behaved NP spin coefficients. Almost inevitably there will be residual
differential equations, which we would hope would be in a reasonably simple
form
in the chosen coordinate system; However,  this often may not be the case, and
we
may then wish to modify our final coordinate choice to make the differential
equations more manageable.

The choice of the four coordinate candidates in step 3 is a crucial, and
difficult step.
Held [6] originally  envisaged   an $`$optimal situation' where
 (i) six real   quantities --- more precisely,  four functionally independent
zero-weighted quantities and one complex weighted quantity, [11] ---
are suggested explicitly by the spin coefficients and Riemann tensor
components, and (ii) the integration procedure, after the first three steps,
 yields only a complete involutive set
of tables of the GHP operators on these six quantities.
 (In such a situation the problem
would be essentially solved; since the differential operators ---
equivalently the tetrad components --- could be written down directly from the
tables with the four zero-weighted quantities as coordinates, and the complex
weighted quantity gauged to unity.)     Unfortunately, so far, in any practical
application of this method,  it has not been possible to
obtain, directly from the spin coefficients and Riemann tetrad components, the
required  four
functionally independent zero-weighted quantities.
 In practice, less than the full quota of four
zero-weighted quantities is usually supplied by the spin coefficients and
Riemann tensor;
at this stage it has been customary [12] to make
a translation  back to the NP formalism  or    to introduce   the `missing
coordinates' from outside the formalism [6,7,10,13].
 So, in fact, usually a compromise has had to be made at  this stage, and  the
ideals of the coordinate-independent integration procedure in
the GHP formalism
as proposed by Held  have had to be modified.  However,  we shall show in this
paper that it is possible, even when the spin coefficients and Riemann tensor
components fail to provide the full quota of zero-weighted quantities, to
continue working in the GHP formalism along the lines proposed by Held.

 We now know
 why, in practice, it has not been possible to obtain
the full quota of four functionally independent zero-weighted quantities from
the spin coefficients and Riemann tensor components:   the investigations
in GHP formalism
which have been  carried out were in specialised classes of spaces, which
almost inevitably
 means that there exists at least one Killing vector.  This in turn
means the existence of at least one cyclic coordinate in the metric, and hence
{\it our inability to obtain four functionally independent quantities from
quantities constructed from this metric} (e.g. from spin coefficients and
Riemann tensor components). However, recently, without going outside the GHP
formalism,   a means of overcoming this problem has been proposed and
illustrated with an example
 in [11].  The space under consideration (implicitly) contained two
Killing vectors, and it was found that the spin coefficients and Riemann tensor
components only supplied {\it two} functionally independent scalars;
the additional two coordinate candidates were obtained by taking zero-weighted
`potentials' for  some of the spin coefficients and Riemann tensor components.
The procedure for picking out such a potential relies on intuition rather than
any standard procedure; we would prefer a more algorithmic method where
various possible choices of coordinate candidates can be generated and tested
directly, and the most suitable chosen.

In this paper we present such an alternative method for choosing
coordinate candidates. Fortunately, in spaces containing Killing vectors, there
is another source from which we may find additional coordinate candidates; if
we   include explicitly the Killing vector equations alongside the other field
equations, then  --- since the Killing
vector components are not constructed from the metric --- {\it we can look to
the
Killing vector components for additional scalars, functionally independent of
the metric}. So we have found another possible way of satisfying the first
condition of Held's optimal situation --- although to obtain the required six
quantities we have  to go beyond the spin coefficients and Riemann tensor
components, which was Held's original proposal. This is precisely the situation
which occurs in the $G_2$ class of the NT spaces, and makes it an ideal
application to illustrate this method. (We should emphasise that  we are
not claiming, in this paper, that the Killing vector components will {\it
always} supply {\it all} the missing coordinate candidates; the relationship of
zero-weighted GHP quantities to Killing and homothetic Killing vectors needs
careful and detailed treatment, which will be given elsewhere. In this paper,
we
are simply pointing out the possibility of using these components, and giving
an
application where it is possible.)

\medskip

It is emphasised that once  the  tables are found for the derivatives of our
four
coordinate candidates and for the one complex weighted quantity, then ---
alongside the Ricci, Bianchi and Killing equations --- we have {\it all}\/ the
information from all the equations.   We would hope that this complete system
would reduce to satisfy  the second condition of
Held's `optimal situation'  i.e. that only a complete involutive set of
six tables remain. However,  this is  unlikely in practice;  we can expect that
in the tables some additional unknown functions will
 occur, which have to satisfy some residual  differential equations
from the Ricci,  Bianchi and Killing sets. In such a situation the choice of
the
coordinate candidates as coordinates is not necessarily the best choice; an
alternative choice may make the residual differential equations more
manageable.
At this stage we could of course  write down the differential equations with
the coordinate candidates as coordinates, and attempt to obtain further
simplifications by explicit coordinate transformations, in the usual manner.
However, we believe that  there are still advantages to be gained by keeping
within the GHP formalism, obtaining further simplification within that
formalism, and finally allowing the structure of the suitably simplified GHP
operator equations to suggest the optimum coordinate choices.
Again,  the application in this
paper illustrates precisely how this procedure can be implemented.

\

{\bf 1.2.  Outline of the  integration procedure: the $G_2 $ case in the
NT problem.}

For the particular problem being considered in this paper the complete GHP
system consisting of the three sets of equations --- Bianchi, Ricci and
commutator equations --- has to be supplemented by three other sets of GHP
equations: the Killing vector equations, the homothetic Killing vector
equations, and the non-Abelian $G_2 $ condition. The five steps in the
integration procedure will then be carried out for the whole system of
six sets of equations. Before beginning, we will of course ensure that all the
equations are specialised to  vacuum Petrov Type $N $ spaces.

 Held, in [8,9], has investigated  some
algebraically special vacuum spaces admitting  Killing vectors, and in Section
2 and in the Appendix  we clarify some differences in approach and notation
between his work and our previous paper, [1]; in Section 3 we
 make use of the first part of Held's
 work in [8] --- the $\rho-$integration ---  and specialise
the relevant calculations  to    Petrov type N spaces, as well as
extending to the $G_2 $ case.
This $\rho-$integration (corresponding to the first two steps of the
integration procedure) is summarised at the beginning of  Section 3, and we
then also   carry out the third step --- applying the commutators to  six
appropriate quantities. The choice of these quantities is dictated by our wish
to keep the calculations as simple and manageable, and to end up with as
concise a system, as possible;  however, we also try to keep in touch with the
earlier method in [1], and to draw comparisons. Since the spaces under
consideration have one Killing vector we will not be able to obtain all our
coordinate candidates directly from the spin coefficients and Riemann tensor
components; in fact  none of our candidates is chosen in this manner:  three
candidates are constructed from the tetrad  components of the Killing vector
and
homothetic Killing vector combined with  the spin coefficients, while the
fourth
candidate is chosen as a potential for a combination of some of these
quantities.
Since we choose our six quantities as far as possible to suit the structure of
the system of equations, and to give as simple a presentation as possible,
when
the tables are constructed for these six  quantities we find  considerable
redundancy and simplifications once these equations are put  alongside the
Ricci, Bianchi and
Killing equations.  Unfortunately, we do not achieve Held's optimal situation,
since in addition to the six quantities, the six tables
 involve other  functions,
which are themselves subject to two residual Ricci equations.

The fourth and fifth steps of the integration procedure involve the choice of
coordinates and gauge. In the last part of Section 3 we make the most
straightforward choice ---  choosing the coordinate candidates as
coordinates. We find that these choices do cause considerable structural
simplification; in particular they enable the  residual partial differential
equations to separate into a pair of coupled
 ordinary differential equations for one  unknown complex function of one real
coordinate. However, these residual ordinary differential equations do not look
very manageable, and, in particular, decoupling seems a problem.

In our work in Section 3 our primary concern is to ensure that we consider a
sufficient system of equations; up to  this stage we are not so concerned
with the most efficient presentation of the residual partial differentiatial
equations. However,  we do keep in mind the need for subsequent simplification,
and of course we are looking for the simplest options; and in
particular we are hoping for
candidates which permit  separation. So although, in this case, we will
ultimately choose
 coordinates different from the coordinate candidates, yet we realise that they
will have to be closely related in order to share the separation property;
specifically we will retain the first three coordinate candidates as
coordinates, but we will allow the structure of the residual pair of ordinary
operator differential equations to suggest our fourth coordinate.

The great advantage of our GHP  approach
is that we can get an overall picture of the structures of the equations, and
this leads to
insights not just into   separability  but also into redundancy, decoupling and
order
reduction  in a manner which does not first require a precise coordinate
choice;
 the final coordinate choice can then be made in an informed manner which best
exploits the separability and decoupling properties, and seeks for maximum
order
reduction and/or simplification in the final differential equation.
These advantages are demonstrated and illustrated in the remaining
sections.

 In
Section 4 we go back to the residual pair of Ricci equations as given in the
GHP
formalism, and by modifying the GHP weighted operator $\tilde \Ph' $ to a new
zero-weighted operator ${\cal D} $ we obtain a simpler version of these
operator equations, where it is explicit that all the terms depend on only one
real variable. We consider this version of these equations as our `master
equations'. From these equations we are able to experiment with various
different approaches to decoupling and reduction of order; in particular, we
find
one particularly structurally simple presentation of the master equations, as a
set of first order ordinary differential operator equations in complex
functions.

In Section 5, we show that by choosing our fourth coordinate from a class which
exploits the structure of this simple version of the master equations, the
decoupling problem is less complicated than in the previous coordinate choice
in
Section 3;
 by a
little manipulation the problem reduces   to one  real third order
 differential operator equation for a complex function of a real variable.
However, we still have the freedom to make our explicit coordinate choice, and
by a suitable choice the problem easily reduces   to one  real third order
differential equation for a real function of a real variable. We note that
there
is a class of such coordinates which reduce the problem to this order.

In Section 6 we present an alternative and more algorithmic approach to
obtaining the residual equation.  By a systematic approach we construct, from
our set of complex first order equations, a  set of {\it real} first
order equations in the form of a closed chain. These equations suggest both the
cordinate and dependent variable in a natural way;  the decoupling problem then
disappears, and we obtain a comparatively concise form for the residual  real
third order differential equation.

\medskip

The early work on the $G_2 $ case of the NT  problem by McIntosh [14]   reduced
the
problem to
 a complicated  third order complex differential equation for a complex
function
of a real variable, which he rearranged to a  sixth order complex differential
equation  for the complex function alone --- without its complex conjugate
occuring  explicitly. In recent years the problem has been reduced to
essentially a single real ordinary differential equation of a real function of
a real variable;
 the order of the final differential equation has gradually been
lowered (at the expense of increasing non-linearity) and most recently there
have been obtained a few  (very complicated non-linear) third
order real differential equations  for a real unknown function, each
highly dependent on its background formalism and particular coordinate choice
[15-18]. The only known solution, the Hauser solution [19,20,21], is found to
be a singular case.  Each of these analyses began with a preferred coordinate
system, and a lot of involved manipulation and complicated coordinate changes
were needed in order to arrive at the final third order equation; it is not
clear
from each individual case whether alternative (and perhaps simpler) third (or
lower) order equations could be obtained, nor is there any obvious links
between
the very different equations so far obtained. On the other hand, our method
supplies us with a whole class of third order equations, and with the
possibility of obtaining even more in a manner where we have some understanding
of the structures involved, as well as some freedom to  simplify these
structures.

We highlight the insights we have obtained into the GHP
integration procedure and into this particular application in the concluding
section; we also outline how   the
various differential equations obtained previously  by other
approaches to this application can be deduced from our master equations.

\

\

{\bf 2. Formalism and Notation.}

As noted above, Held [8,9] has used the GHP formalism to investigate
algebraically
special vacuum spacetimes admitting Killing vectors, and we shall make direct
use of some of his results in this work. Some points need to be clarified at
the
outset:

\medskip
Firstly,
Held, [6,8,9]  slightly modifies three of the usual GHP operators $ \Ph', \Ed,
\Ed'
$ to $ \tilde \Ph', \tilde \Ed, \tilde \Ed'  $; this makes no essential
difference to the overall structures of the GHP formalism, but simplifies
calculations by taking advantage
of some  properties of algebraically special vacuum metrics. In
particular, it is easy to carry out a $`$coordinate-free $\rho$-integration'.

\medskip
Secondly, as noted in Section 1,  the set of Killing equations with
which Held [8,9] works differ a little --- being somewhat simpler --- from the
conventional set, and in particular from the set used in [12], and in the
previous paper. We show in  Appendix I the relationship between these two sets,
and confirm that the two apparently different sets of equations in the two
papers are equivalent for Killing vectors. We also generalise the set in [8,9]
for the presence of homothetic Killing vectors, and show its equivalence to the
analogous set in [12].

  \medskip Thirdly, there is a question of  notations, and
this needs to be set  out in detail to avoid confusion:

\smallskip
(i) In the previous paper, [1] we used the symbols  $b,\bar c, a $ for tetrad
components of Killing vectors, and
$\Sigma^o $ for the twist. Held uses the symbols
$\xi_0, \xi_1,\xi_2 $, [9] or $\theta_0,\theta_1,\theta_2 $, [8] for Killing
vector
components, and
$\Omega^o (=-2i\Sigma^o) $ for the twist. When we make use of Held's results we
shall translate them into the former symbols for consistency with the previous
paper.

\smallskip
(ii) Quantities which are annihilated by the operator
$\Ph $ are labelled in Held's work with \ ${}^o $ \ e.g. $\Ph \eta^o=0 $; an
equivalent notation has already been adopted in the previous paper.

\smallskip
(iii) Standard GHP usage, also followed by Held, uses the prime notation for
 half of the spin coefficients, and half of the GHP operators; in this paper we
shall also use that notation.
Although our previous paper uses  the more familiar NP versions of the spin
coefficients, since only $\kappa' (=-\nu )$ and
 $\rho' (=-\mu)$ occur explicitly in this paper there
will be no difficulty in comparison.

\smallskip
(iv) The letter $P $ has a  special usage in  NP notation [3,4], and has been
used extensively in the previous paper, [1];  the related $P^o $ plays an
important role in Held's work [8,9,10], and we retain it in this paper. If  a
 comparison is being made we must take into account a factor of
2, ($P^o\sim 2P $) as can be seen in the defining equations in each paper.
 We also point out that $\tilde P $  defined in [1] is real.

\

\

{\bf  3. An operator-integration approach with coordinate candidates
as coordinates.}

{\bf 3.1. Steps 1,2: The $\rho-$integration.}

 We specialise the $\rho-$integration results in [8]  to   the Petrov type N
case with the substitutions  $$
\Psi_2=0=\Psi_3 \eqno(3.1 )$$
giving:

\

{\it The residual Bianchi and Ricci equations.}

$$
\eqalign {\rho'  = \bar\rho{\rho'}^o}
\eqno(3.2a)
$$
$$
\kappa'  =  {\kappa'}^o
\eqno(3.2b)
$$
$$\Psi_4  =  \rho\Psi_4^o
\eqno(3.3 )$$
with
$$
\eqalign {\Psi_4^o  = -\tilde \Ed'{\kappa'}^o}\eqno(3.4a )$$
$$
\tilde \Ed'{\rho'}^o  =  2i\Sigma^o{\kappa'}^o
\eqno(3.4b)$$
$$
\tilde \Ed{\kappa'}^o  =  \tilde \Ph'{\rho'}^o
\eqno(3.4c )$$
$$
\tilde \Ed\tilde \Ed'\Sigma^o  =  2\Sigma^o\bar\rho'^o
\eqno(3.4d )$$
and
$$
\Ph \rho=\rho^2
\qquad\qquad\qquad
\tilde \Ph' \rho=\rho^2\bar\rho'^o
$$
$$
\tilde \Ed \rho=0
\qquad\qquad\qquad
\tilde \Ed' \rho=-2i\rho^2\tilde \Ed'\Sigma^o
\eqno(3.5 )$$

\

{\it The commutator equations.}

$$
\eqalign {[\Ph,\tilde
{\mathord{\mkern5mu\mathaccent"7020{\mkern-5mu\partial}}}] & =  0}\eqno(3.6a)$$
$$
[\Ph,\tilde
{\mathord{\mkern5mu\mathaccent"7020{\mkern-5mu\partial}}}']  = 0
\eqno(3.6b)$$
$$
[\Ph,\tilde\Ph']   =  0
\eqno(3.6c)$$
$$
[\tilde \Ph',\tilde
{\mathord{\mkern5mu\mathaccent"7020{\mkern-5mu\partial}}}]  =
-{\bar\kappa'^o\over \bar \rho}\Ph +q\bar\kappa'^o
\eqno(3.6d)$$
$$
[\tilde \Ph',\tilde
\Ed']  =
-{ {\kappa'}^o\over  \rho}\Ph +p{\kappa'}^o
\eqno(3.6e)$$
$$
[\tilde \Ed,\tilde \Ed']  = ({\bar \rho'^o \over  \rho}-{\rho'^o \over
\bar \rho})\Ph
-2i\Sigma ^o\tilde\Ph'+p{\rho'}^o -q \bar\rho'^o\eqno(3.6f)
$$
(In addition to the Bianchi and Ricci equations given above there are the two
 equations  $$\eqalign {
2i\tilde \Ph'\Sigma^o & =  -\bar\rho'^o+\rho'^o\cr
\tilde  \Ed \Psi_4^o & = 0}
$$ which follow  from [8]. However,  both are identically satisfied by virtue
 of the other Bianchi and Ricci equations, given above, and the
commmutators. We show this as follows:  apply commutator (3.6f) to $\Sigma^o$
and
use (3.4d)  and its conjugate to get the first;
apply commutator (3.6f) to $\kappa'^o$, followed by (3.6e) applied to
$\rho'^o $ and
then use appropriate equations from (3.4) together with the first equation to
get
the second.)

\

{\it The equations for one Killing vector.}

When the   Killing vector $
\xi_1$ is given by
$$
\xi_1^{\mu} = a_1l^{\mu}+b_1n^{\mu}-c_1 m^{\mu}-\bar c_1\bar m^{\mu}
\eqno(3.7 ) $$
the  Killing equations
can be $\rho$-integrated  to obtain, [8] $$
\eqalign{a_1 & =  -2i\bar c_1^o\tilde
\Ed'\Sigma ^o-\bar\rho'^o b_1^o+i\Sigma ^o\tilde \Ph'b_1^o +
{1\over 2} ({1\over \rho}+{1\over \bar \rho})\tilde \Ph'b_1^o
\cr
b_1 & =  b_1^o
\cr
c_1 & =  c_1^o/\bar \rho}
\eqno(3.8)$$
with
$$
\eqalign {\tilde \Ph'b_1^o  =
-{1\over 2}(\tilde \Ed c_1^o+\tilde \Ed' \bar c_1
^o)}
\eqno(3.9a )$$
$$
\tilde \Ed b_1^o  =  -2i\Sigma^o \bar c_1^o
\eqno(3.9b )$$
$$
\tilde \Ph'c_1^o  =  \ 0
=
\tilde \Ed' c_1^o\eqno(3.9c )$$

$$
\eqalign {
b_1^o\tilde \Ph'\rho'^o-\bar c_1^o\tilde \Ed'\rho'^o- c_1^o\tilde
\Ed\rho'^o +2\rho'^o\tilde \Ph'b_1^o & =  0}\eqno(3.9d )$$
$$
b_1^o(\bar\rho'^o-\rho'^o) +2ic_1^o\tilde \Ed\Sigma ^o+2i\bar c_1^o\tilde \Ed'
\Sigma
^o +i\Sigma^o (\tilde \Ed c_1^o+\tilde \Ed'\bar c_1^o)  =  0\eqno(3.9e)$$
where the subscript \ ${}_1 $ \   denotes quantities associated with the
 Killing
vector $\xi_1 $.

The last equation was not explicitly displayed in [8]  but is given in [9]
where it is pointed out that it follows from the condition that $a_1 $ is real.

However,  when we examine this last equation  we note that it is simply the
commutator (3.6f) applied to $b_1^o $. In addition, when we apply the
commutator
(3.6d) to  $b_1^o $ and differentiate the resulting equation by $\tilde \Ed' $,
after using (3.6f) a number of times together with some equations from (3.4),
we
obtain (3.9d).  Hence the two equations (3.9d,e) are  identities modulo  the
residual
equations and the commutators, and so can be omitted.

\

{\it The equations for one homothetic Killing vector.}

When the homothetic Killing
vector $\xi_2 $ is given by
$$
\xi^{\mu}_2 = a_2l^{\mu}+b_2n^{\mu}-c_2 m^{\mu}-\bar c_2\bar m^{\mu}
\eqno(3.10 ) $$
we  use the set of equations  (AI.2) in the first Appendix, and in the
same manner as in [8] we can easily integrate  to obtain
 $$
\eqalign{a_2 & = -2i\bar c_2^o\tilde
\Ed'\Sigma ^o-\bar\rho'^o b_2^o+i\Sigma ^o\tilde \Ph'b_2^o +
{1\over 2} ({1\over \rho}+{1\over \bar \rho})(\tilde \Ph'b_2^o
-\phi)\cr
b_2 & = b_2^o
\cr
c_2 & = c_2^o/\bar \rho
}\eqno(3.11 )$$
with
$$
\eqalign{\tilde \Ph'b_2^o  = -{1\over 2}(\tilde \Ed c_2^o+
\tilde \Ed' \bar c_2^o-\phi)}
\eqno(3.12a )$$
$$
\tilde \Ed b_2^o  =-2i\Sigma^o \bar c_2^o
\eqno(3.12b )$$
$$
\tilde \Ph'c_2^o  =  0
=
\tilde \Ed' c_2^o
\eqno(3.12c )$$
$$
\eqalign{b_2^o\tilde \Ph'\rho'^o-\bar c_2^o\tilde \Ed'\rho'^o- c_2^o\tilde
\Ed\rho'^o +2\rho'^o\tilde \Ph'b_2^o  =0} \eqno(3.12d )$$
$$
b_2^o(\bar\rho'^o-\rho'^o)+2ic_2^o\tilde \Ed\Sigma ^o+2i\bar c_2^o\tilde
\Ed'\Sigma ^o
+i\Sigma ^o(\tilde \Ed c_2^o+\tilde \Ed'\bar c_2^o-\phi)  =0\eqno(3.12e)$$

where the subscript \ ${}_2 $ \   denotes quantities associated with the
homothetic Killing
vector $\xi_2 $.
We note that in this case also, the last two equations (3.12d,e) are
identities modulo
the residual equations and the commutators, and so can be omitted.

\

{\it The non-Abelian $G_2 $ condition.}

We now consider the case  for a non-Abelian $G_2 $ of homothetic motions;
by an appropriate choice of basis in the Lie algebra we get
$$
[{\boldxi_1},{\boldxi_2}]+2{\boldxi_1}=0
\eqno(3.13 )$$
where $\boldxi_1(=\xi_1^{\mu}\nabla_{\mu})$
and
$\boldxi_2(=\xi_2^{\mu}\nabla_{\mu})
 $ are the
Lie derivative operators associated with the Killing vector $\xi_1$ and the
homothetic Killing vector
$\xi_2 $ respectively.
Equation (3.13) has to be
satisfied  when applied to an arbitrary
(scalar) quantity  or
equivalently  to four  real functionally independent quantities. In the present
context that means that it must be applied to four real zero-weighted
functionally independent quantities, and {\it for such zero-weighted scalars}
the
operators can be written as
$$\eqalign{{\boldxi_1}(\equiv \xi_1^i\nabla_i) & =a_1\Ph+b_1^o\tilde
\Ph'-c_1^o\tilde \Ed- \bar {c_1}^o
\tilde \Ed'\cr {\boldxi_2}(\equiv
\xi^i_2\nabla_i) & =a_2\Ph+b^o_2\tilde \Ph'-c^o_2\tilde \Ed-\bar c^o_2
\tilde \Ed'}\eqno(3.14 )$$

\medskip

So for the case of Petrov type N vacuum spaces admitting one ordinary
Killing vector $\xi_1 $, and one homothetic Killing vector $\xi_2 $, we would
be
required to solve the residual Ricci and Bianchi equations (3.4,5) together
with
the residual Killing equations (3.9a,b,c),
the residual homothetic Killing equations (3.12a,b,c), the commutator equations
(3.6) and non-Abelian condition (3.13).

\

{\bf 3.2.  Step 3: Choosing the coordinate candidates
and applying the commutators.}

Since the first two steps of the procedure set out in [11] have been carried
out,  the next step
is to ensure that the commutator equations are completely
satisfied;   to ensure this we must apply them to         four
real, zero-weighted, functionally independent
quantities  and to one complex weighted quantity. We have
just noted the similar requirements for (3.13) so it will be convenient to
apply
both sets of equations --- (3.6) and (3.13) respectively --- to the same four
coordinate  candidates.  We would
prefer to select  those four
coordinate  candidates in a manner  which would ensure that any additional
constraint
equations resulting from the commutator equations
assume as simple a form as possible; but we also wish at this stage not to move
too
far from the coordinate choices of the first method. Fortunately, for the first
three
coordinate candidates, these two wishes coincide.

For each of the four real and one complex  quantities we will obtain a table
for the action of the
operators; we may also obtain some new differential equations when the
commutators and $G_2 $ condition are used on each of these quantities, as well
as some
simplifications. We now set out these results systematically for each
quantity in turn.

\

{\it Choice of the first two real (one complex) coordinate candidates, $\zeta
$.}

 Noting the simple equations for the (0,2) quantities
$c_1^o
$ and
$c_2^o
$ we begin with the comparatively obvious  choice  for our first (complex
zero-weighted)
coordinate candidates $$\zeta = i({c_2^o/
c_1^o})\eqno(3.15 )$$ so that
$$ \Ph \zeta = \tilde \Ph'\zeta =\tilde \Ed'\zeta = 0\eqno(3.16 )
$$
(We choose the factor $i $ in our definition of $\zeta  $ in order to make
easy direct comparison with [1], and also because  this
choice  simplifies a later part of the calculation. We also  omit \ ${}^o$ \ on
$\zeta$ , even though it is annihilated by $\Ph$. )

When the commutators are applied to this zero-weighted  quantity the only
non-trivial results are, $$  \Ph P^o =\tilde \Ph'P^o=\tilde \Ed'P^o=0\eqno(3.17
)
$$
where $P^o $ is the (0,-2) quantity,
$$
P^o=\tilde \Ed \zeta
\eqno(3.18 ) $$
Noting the similarity between these equations and those for $c_1^o $, we can
write  $$ \Ph(c_1^o P^o)= \tilde \Ph'(c_1^o P^o) =\tilde \Ed'(c_1^o
P^o)=0\eqno(3.19 ) $$
Since $(c_1^o P^o) $ is also a zero-weighted quantity comparison between its
equations (3.19) and those for $\zeta  $ show that
$$
c_1^o P^o=2g(\zeta )
\eqno(3.20 )
$$
where $g $ is an arbitrary function of $\zeta  $. (The factor $2 $ has been
introduced to give correspondence with  [1])

\medskip
Turning next to equation (3.13), and applying (3.14) to $U$ we find
$$
\eqalign{{\boldxi_1} \zeta   & = -c_1^o \tilde  \Ed \zeta  = -2g(\zeta )
\cr
{\boldxi_2} \zeta & = -c_2^o \tilde  \Ed \zeta =i c_1^o \zeta  \tilde  \Ed
\zeta  = 2i\zeta  g(\zeta ) }\eqno(3.21 )
$$
and so obtain from (3.13),
$$
g(\zeta )=i
\eqno(3.22 )
$$
or
$$
c_1^o={2i\over P^o}\qquad \hbox{and} \qquad c_2^o={ 2\zeta \over P^o}
\eqno(3.23 )
$$
where $\zeta  $ satisfies (3.16) and $P^o  $ satisfies
(3.17).
\medskip
In summary, we have a table  (3.16,18) for the action of the operators on the
two
real coordinate candidates $\zeta  $; and the action of the commutators on
these quantities yields the additional  equations (3.17) ---  the
 table giving
the action of most of the operators on the weighted quantity $P^o $. The result
of
applying the  condition (3.13) has been to obtain explicit expressions for the
Killing vector components $c_1^o,c_2^o. $
\smallskip
(Although this is
not immediately apparent from the respective definitions, a little work
confirms that $\zeta  $ as
defined above essentially agrees with $\zeta  $ as defined in [1]. We could
have chosen to define
$\zeta  $ in this paper in a manner more closely analogous to [1] by
introducing it as a potential
for
$P^o$, which, in turn, could have been introduced as a potential for some of
the spin coefficients.
However,  as we have emphasised before, we wish in this paper to illustrate the
{\it direct} method
of choosing coordinate candidates.)

\

{\it Choice of the third coordinate candidate, $R$.}

A rearrangement of  (3.5)
 $$
\eqalign{\Ph \Bigl(({1\over \rho}+ {1\over \bar
\rho})/|P^o|\Bigr) & = - 2/|P^o|\cr
\tilde \Ph' \Bigl(({1\over \rho}+ {1\over \bar
\rho})/|P^o|\Bigr) & = -(\rho'^o+\bar\rho'^o)/|P^o|
\cr
\tilde \Ed \Bigl(({1\over \rho}+ {1\over \bar
\rho})/|P^o|\Bigr) & = -2i{\tilde \Ed \Sigma^o\over |P^o|}-{\tilde \Ed P^o\over
2
P^o}({1\over \rho}+ {1\over \bar \rho})/|P^o|
\cr
\tilde \Ed' \Bigl(({1\over \rho}+ {1\over \bar
\rho})/|P^o|\Bigr) & = 2i{\tilde \Ed'\Sigma^o\over |P^o|}-{\tilde \Ed'\bar
P^o\over
2\bar P^o}({1\over \rho}+ {1\over \bar
\rho})/|P^o|}\eqno(3.24 ) $$
 suggests an obvious choice for
our third coordinate candidate $R $,
$$R= -({1\over \rho}+ {1\over \bar
\rho})/|P^o| \eqno(3.25)$$
satisfying
 $$
\eqalign{\Ph R & = 2 /|P^o|\cr
\tilde \Ph' R & = (\rho'^o+\bar\rho'^o)/|P^o|
\cr
\tilde \Ed R & = 2i{\tilde \Ed \Sigma^o\over |P^o|}-R{\tilde \Ed P^o\over
{2P^o}}
\cr
\tilde \Ed' R & = -2i{\tilde \Ed'\Sigma^o\over |P^o|}-R{\tilde \Ed'\bar
P^o\over
{2\bar P^o}}}\eqno(3.26 ) $$

 From their respective differential equations given in tables (3.16) and
(3.26),
the three coordinate candidates $ (\zeta +\bar \zeta ), i(\zeta -\bar \zeta ),
R  $ are easily seen to be functionally independent of each other; and when the
commutators are applied to $R, $ the only two commutators (3.6d,f)  which are
not
identically satisfied, give respectively, $$\eqalign{ \bar\kappa'^o  = -\tilde
\Ph'({\tilde \Ed P^o\over {2P^o}})}\eqno(3.27a)$$
$$
\bar\rho'^o  =-\tilde \Ed'({\tilde \Ed P^o\over {2P^o}})
\eqno(3.27b)$$
Substituting these values into (3.4a) shows that
$$\Psi_4^o  = \tilde \Ed'\tilde \Ph'({\tilde \Ed' \bar P^o\over {2\bar P^o}})
\eqno(3.28 )
$$
  and also that (3.4c)
is identically satisfied by virtue of (3.6d).

 So the only remaining Ricci and Bianchi equations to be solved are (3.4b) and
(3.4d) which now become
$$
\eqalign {\tilde \Ed'  \tilde \Ed({\tilde \Ed' \bar  P^o\over {\bar P^o}})   &
=
2i\Sigma^o \tilde \Ph'({\tilde \Ed' \bar P^o\over {\bar P^o}})
\cr
\tilde \Ed\tilde \Ed'\Sigma^o & =  -\Sigma^o\tilde \Ed'({\tilde \Ed P^o\over
{P^o}}) }\eqno(3.29)
$$
 The first of these may be rearranged to
$$
\eqalign {\tilde \Ed  \tilde \Ed'({\tilde \Ed' \bar  P^o\over {\bar
{P^o}^{3/2}}})   & = 0 }\eqno(3.30)
$$
There are also the inequalities
$$
\Sigma^o\ne 0\eqno(3.31a)$$
$$
\tilde \Ed \tilde \Ph'({\tilde \Ed P^o\over {2P^o}})\ne 0\eqno(3.31b)$$

The remaining constraint equations
from
the Killing equations  and from the homothetic Killing equations, after
the substitutions (3.23), are  $$
\eqalign{\tilde \Ph'b_1^o & = i({\tilde \Ed P^o\over {{P^o}^2}}-{\tilde \Ed'
\bar P^o\over
 {\bar {P^o}^2}})\cr
\tilde \Ed b_1^o & =-4\Sigma ^o/{\bar P^o}\cr
\tilde \Ed b_2^o & =-4i\bar \zeta \Sigma ^o/{\bar P^o}\cr
\tilde \Ph'b_2^o & = -2+{1\over 2}\phi+{\zeta \tilde \Ed P^o\over {{P^o}^2}}+
{\bar \zeta \tilde \Ed' \bar P^o\over
 {\bar {P^o}^2}}}\eqno(3.32 )
$$

\medskip
Turning next to equation (3.13), and applying (3.14) to $R $ we find
$$
\eqalign{{\boldxi_1} R   & = 0
\cr
{\boldxi_2} R & =  (2+{\phi  \over 2 })R}\eqno(3.33 )
$$
 and so we easily confirm that (3.13) is satisfied identically when applied to
$R
$.

\medskip
In summary, we have a table  (3.26) for the action of the operators on the
coordinate candidate $R  $; and the action of the commutators on $R $
 yields explicit expressions for some of the spin coefficients and $\Psi_4^o
$; some of the Killing equations are also simplified. The result of applying
the
 condition (3.13) to $R $ yields no new information.

\

{\it Choice of the fourth  coordinate candidate, $u$.}

 We have noted in the introduction to this paper that we can
introduce each of the coordinate candidates  either directly as one of the
(combination of) existing elements in the GHP formalism, or as a $`$potential'
for
 existing elements.  Such potentials are chosen by a
careful consideration of the structure of the equations, especially the
commutators, paying particular attention to the appropriate weights, [11];
specifically the four derivatives $ \Ph \eta, \   \tilde \Ph' \eta,
\ \tilde   \Ed \eta, \
\tilde \Ed' \eta\ $ of a zero-weighted potential $ \eta $ are equated to
(combinations of) some of the spin coefficients and Riemann tensor components
in
such a way that the commutators acting on $\eta$ are identically satisfied.
   Since for
zero-weighted $ \eta $,
$$\eqalign{
\nabla_{\mu}  \eta =n_{\mu}\Ph\eta + l_{\mu} \tilde\Ph'\eta -\bar m_{\mu}
 \tilde\Ed\eta
- m_{\mu}  \tilde\Ed'\eta}\eqno(3.34)$$
clearly fixing its GHP derivatives  determines $\eta $ uniquely (up
to an addititive constant).

 \smallskip
  We can look for hints
for possible potentials by rearranging the residual equations, paying
particular attention to weights; for instance the first and last of (3.32) can
be combined to give  $$\tilde  \Ph'(b_2^o-ib_1^o\bar
\zeta )=(\bar \zeta + \zeta ){\tilde \Ed P^o\over {{P^o}^2}}-{{4-\phi}\over 2}
\eqno(3.35 ) $$
i.e.
$$ \eqalign{\tilde \Ph' & \Bigl({P^o|P^o|(b_2^o-ib_1^o  \bar \zeta
)\over ( \bar \zeta + \zeta )}\Bigr)  =2{\tilde \Ed
|P^o|}-{(4-\phi)P^o|P^o| \over 2( \bar \zeta + \zeta )}\cr
  & =2(\bar \zeta + \zeta
)^{1+m}\tilde \Ed \Bigl(|P^o|(\bar \zeta + \zeta
)^{-1-m}\Bigl)\quad\quad\hbox{with} \quad m=-{\phi  \over 4 }}\eqno(3.36
)$$

Therefore, the choice
$$
\Ph u = 0\eqno(3.37a )$$
$$\eqalign{\tilde \Ph'u =  |P^o|(\bar \zeta +\zeta
)^{-1-m}}\eqno(3.37b )$$
$$ \tilde \Ed u =   P^o|P^o|(b_2^o-ib_1^o\bar
\zeta ) ( \bar \zeta + \zeta )^{-2-m}/2
\eqno(3.37c )
$$
where  $u $ is a real zero-weighted quantity,
guarantees that  commutator (3.6a)   is
satisfied identically when applied to $u $. (In fact it follows easily that
 (3.6f) is the only commutator not satisfied identically when applied to $u$.)
\medskip
We now need to check explicitly that the four coordinate candidates
$R,\zeta,\bar \zeta,  u, $ are
functionally independent.  We can write the four derivatives of each of these
four real zero-weighted quantities in determinant form as
$$\left|\matrix{
 2 /|P^o| &
 ( \rho'^o+\bar \rho'^o)/{|P^o|} &
 {(2i\tilde \Ed\Sigma^o/ {|P^o|})}-R\tilde \Ed P^o/ 2 P^o
 &  (-2i\tilde \Ed'\Sigma^o/ {|P^o|})-R\tilde \Ed'\bar P^o/
2\bar P^o\cr 0&0&P^o&0\cr
0&0&0&\bar P^o\cr
0&
 |P^o|(\bar \zeta +\zeta
)^{-1-m} & P^o\ell ( \bar \zeta + \zeta )^{-1}&\bar
P^o\bar \ell ( \bar \zeta + \zeta )^{-1}
}\right|\eqno(3.38 )
$$
where we have defined the function $\ell(\zeta ,\bar \zeta , u) $
by,
$$\eqalign{
\ell ={{(\zeta +\bar \zeta )\tilde \Ed u}\over {P^o}}}\eqno(3.39)$$
Hence, using (3.39), (3.6d) and (3.37b), we have
$$
{\tilde \Ed P^o \over {P^o}^2}=2{\tilde \Ph'\ell\over |P^o|}
(\zeta +\bar \zeta )^{m}+2(1+m)(\zeta +\bar \zeta )^{-1}\eqno(3.40)
$$
It is easy to confirm that this determinant   is non-zero.  (It can be seen
that this definition for $\ell$ agrees with the definition in the previous
paper,
[1].)

 \medskip

When we apply the remaining commutator (3.6f) to $u $   we obtain
$$
\eqalign{2|P^o|\Sigma^o  & =i( \tilde \Ed  \tilde \Ed' u  -
\tilde \Ed'  \tilde \Ed u)(\bar \zeta
+\zeta )^{1+m}
}\eqno(3.41)$$

which becomes
$$
\eqalign{2|P^o|{\Sigma^o } & =i\Bigl( \bar P^o \tilde \Ed \bar \ell
 -
P^o \tilde \Ed' \ell+|P^o|^2(\bar \zeta
+\zeta )^{-1}( \ell-\bar \ell)\Bigr)(\bar \zeta
+\zeta )^{m}
}\eqno(3.42)
$$

We can find  $b_1^o,b_2^o $ explicitly from (3.37c),
$$
b_1^o= 2 i { {( \ell-    \bar \ell
 )(\zeta +\bar \zeta )^m}\over |P^o|}
\eqno(3.43)
$$
$$
b^o_2= 2 {{(\zeta \ell+
\bar \zeta   \bar \ell)(\zeta +\bar \zeta )^m}\over |P^o|}
\eqno(3.44 )
$$
and when we substitute these values into the remaining Killing
equations (3.32)  we get
$$
{\tilde \Ed  \ell \over P^o}= \ell\,{\tilde \Ph'
\ell \over |P^o|}
(\zeta +\bar \zeta )^{m} \eqno(3.45a ) $$
$${\tilde \Ed'   \ell \over \bar P^o}= \bar \ell\,{\tilde \Ph'
\ell \over |P^o|}
(\zeta +\bar \zeta )^{m} \eqno(3.45b )
 $$
We can now use (3.39) and (3.37b) to rewrite (3.37c) as
$$ {\tilde \Ed u \over P^o} =  \ell
{\tilde \Ph' u\over |P^o|}( \bar \zeta + \zeta )^m \eqno(3.46)
$$
and a comparison with (3.45) shows that these equations give  the very
crucial simplification that $\ell$ and
 $u$ are functionally dependent.

When (3.45) are substituted into (3.42) we obtain
$$
\eqalign{2{\Sigma^o\over |P^o|} & =i\Bigl( {\ell\,\tilde \Ph'  \bar \ell
\over |P^o|} -
{\bar \ell\, \tilde \Ph'  \ell\over  |P^o|}+(\bar \zeta
+\zeta )^{-1-m}( \ell-\bar \ell)\Bigr)(\bar \zeta
+\zeta )^{2m}
}\eqno(3.47 )
$$

The two residual Ricci and Bianchi equations (3.29) now become  equations for
$\ell$ by the substitutions (3.40,47).
We now seem to have too many  equations --- one real  and one
complex   --- for the complex unknown $\ell $. However, we shall see that this
is  compensated for by the fact that these two equations  are not
independent; it will be easier to show this explicitly when we introduce our
coordinates.

The inequalities (3.31) can also be  rearranged with these substitutions.

\

Turning next to equation (3.13), and noting that
$$
\eqalign{ {\boldxi_1} u   & = 0
\cr
 {\boldxi_2} u & = 0 }\eqno(3.48 )
$$
we confirm that (3.13) is satisfied identically when applied to $u $.

\medskip
In summary, we have a table  (3.37) for the action of the operators on the
coordinate candidate $u  $; and the action of the commutators on
$u $  gives
explicit expressions in terms of $\ell$ for the twist $\Sigma^o $ and for the
Killing vector
components
$b_1^o,b_2^o $, as well as the action of the remaining operator $\tilde \Ed $
on
weighted $P^o $. The remaining two Killing equations   give the important
result
that the complex function $\ell$ and the coordinate candidate $u$ are
functionally dependent.  The result of applying the  condition (3.13) to $u $
yields no new information.

\

{\it Choice of one complex weighted quantity, $P^o$.}

When we apply the commutators to $P^o $  and  all the earlier equations are
taken into account, all the commutator equations acting on $P^o$ are
identically
satisfied.

\medskip

{\it Summary of results in this subsection.}

Having confirmed that our  four
coordinate candidates are functionally independent, we now know that we have
extracted all the information available from the commutators and the
non-Abelian
condition.    The net result  is that we have obtained explicit expressions for
all the spin coefficients, Weyl tensor component and Killing vectors
components,
as well as six tables of twenty-four (real)  equations --- (3.16,18) and
their complex conjugates, (3.24), (3.37a,b,39), (3.17,40) ---
for the action
of the  operators  on the four coordinate candidates $R, u,
\zeta , \bar \zeta  $ and on the one complex weighted
quantity, $P^o $. However,  this set of tables is not self-contained; it
contains explicit
expressions in $\ell$, which is a function of $u $. The quantity
$\ell $ must satisfy the two residual  Ricci equations   (3.29),
subject to the inequalities (3.31), when the substitutions (3.40,47) are made.

  \smallskip
The next obvious step is
to use   the
coordinate candidates as coordinates.

\

{\bf 3.3.  Steps 4,5:  Choice of  coordinate candidates $R,  \zeta, \bar
\zeta, u  $ as coordinates
and choice of gauge.}

We will now choose the coordinate candidates as coordinates $R,  \zeta, \bar
\zeta, u  $,  which means that their respective  four
tables of  equations (3.16,18),  (3.24) and
(3.37)  become identities, simply defining, in these coordinates, the GHP
operators {\it when acting on zero-weighted quantities } as,
$$\eqalign{
\Ph \equiv \ &\  \Ph (R)\,\partial_R +\Ph(\zeta)\, \partial_\zeta +\Ph(\bar
\zeta)\, \partial_{\bar  \zeta}+\Ph (u)\, \partial_u
= 2|P^o|^{-1} \partial_R
\cr
 \tilde \Ph'= &  ( \rho'^o+\bar \rho'^o)|P^o|^{-1} \partial_R+|P^o|(\bar
\zeta +\zeta )^{-1-m}\partial _u
\cr
 \tilde \Ed= & \Bigl(2i{\tilde \Ed\Sigma^o\over |P^o|}-R{P^o \tilde
\Ph'\ell\over |P^o|}(\zeta +\bar \zeta )^m-RP^o(1+m)(\zeta +\bar \zeta
)^{-1}\Bigr)\partial_R +P^o\partial _\zeta  +P^o\ell (\bar \zeta +\zeta
)^{-1}\partial _u  \cr
\tilde \Ed'=  & \Bigl(-2i{\tilde \Ed'\Sigma^o\over |P^o|}- R{\bar P^o \tilde
\Ph'\bar \ell\over |P^o|}(\zeta +\bar \zeta )^m-R \bar P^o (1+m)(\zeta +\bar
\zeta )^{-1}\Bigr)
\partial_R+\bar P^o\partial _{\bar
\zeta } +\bar P^o\bar \ell (\bar \zeta +\zeta )^{-1}\partial _u  }\eqno(3.49)$$

\medskip
The Killing operator becomes
$$\eqalign{
  \boldxi_1= -2i\partial_\zeta +2i\partial_{\bar \zeta }}\eqno(3.50)$$
while the homothetic Killing operator becomes
$$\eqalign{
  \boldxi_2= -2\zeta \partial_\zeta -2\bar \zeta \partial_{\bar \zeta
}-4mR\partial_R}\eqno(3.51)$$

There is of course also the  table of operator equations  for  the
weighted $P^o$, but this simply  yields the badly behaved spin coefficients
(if we want them) when we make our choice of gauge.

\medskip
This coordinate choice exploits the separability property, and we find that
 the problem has been reduced   to  only two differential equations which can
now be
written out explicitly in these coordinates.
Since $\ell$ is a
function of $u $ only, i.e $\ell \equiv \ell(u)$,
 we can replace  $\tilde \Ph' \ell $   with
$$\tilde \Ph' \ell = \dot
\ell \, \tilde \Ph' u = \dot
\ell |P^o| (\bar \zeta +\zeta )^{-1-m}\eqno(3.52)$$
where \  $\dot {} $\   denotes differentiation with respect to $u $
and we have used (3.37). So the only
remaining equations to be solved are the two residual Ricci and Bianchi
equations (3.29)
for  $\ell $ which is a complex function of the real coordinate $u$. When the
substitutions (3.42,47)  together with (3.58) are made into the equations
(3.29) and
the inequalities (3.31)
 we obtain
$$
\eqalign { & \bar \ell \ell  \  \DDD{ \ell}+3\bar \ell \dot {
\ell}
\ddot { \ell}+\bar \ell(2m+1) \ddot { \ell}-2 \ell {\ddot { \ell}}-2
{\dot{  \ell}}{}^2-(4m+2) \dot{  \ell}-2m(m+1)=0}\eqno(3.53)$$
$$\eqalign{ & \qquad \{\ell{\bar \ell}^2 \   \DDD \ell+  \bar \ell
\ddot \ell\Bigl((2m+1)\bar \ell+(2m-3)\ell+  3\bar \ell\dot \ell+\ell
\dot {\bar \ell}\Bigr) + \bar \ell\dot \ell^2  \dot {\bar \ell}+(2m-3)\bar \ell
\dot \ell^2  \cr & \qquad + (2m+1)\bar \ell \dot \ell   \dot {\bar \ell}
+ (4m-2)\bar \ell \dot {\bar \ell}-(4m^2-4m-3) \ell \dot {\bar \ell}
-(4m^2-4m-2) \ell \}\cr
& \qquad - \{\hbox{c.c.}\}\cr
& = \ 0}
   \eqno(3.54)$$
$$\eqalign{\bar \ell \dot \ell- \ell \dot{\bar \ell}+
\bar \ell- \ell \ne 0
}\eqno(3.55)$$
$$\eqalign{\ell \DDD\ell+3\ddot\ell \dot \ell +(2m+1)\ddot \ell \ne 0
}\eqno(3.56)$$

(These correspond to the
equations and inequalities (111,112) in the previous paper [1],
as well as  to the  equations and inequalities
  (25,26,28,29) originally given in [17], when the latter two sets of equations
are specialised to $f=$ constant.)
\smallskip

We  can substitute (3.53) (and its complex conjugate) into (3.54)
to obtain a simpler version,
$$\eqalign{ &  \{\bar \ell \ell      (   \dot{\bar
\ell}        +2m-1) \ddot \ell       +\bar
\ell{ \dot\ell}^2\dot{\bar \ell}       + (2m-1)\bar  \ell{
\dot\ell}^2      \cr & \quad +(2m+1)\bar \ell \dot\ell\dot{\bar
\ell}      + (4m^2-1)\bar \ell{ \dot\ell}       -(4m-2) \ell {
\dot\ell}       -(6m^2-2m-2) \ell \} \cr & \ \
 -\{\hbox{c.c}\} \cr & \  = 0 }
\eqno(3.57)$$
We now  find explicitly the redundancy relation  between this pair of equations
to be  $$\eqalign{ {d\over du} \bigl[\hbox{eq.}(3.57)\bigr]= (2m-1+  \dot {\bar
\ell}) {\bigl[\hbox{eq.}(3.53)\bigr]}-  (2m-1+        \dot \ell)
\overline {\bigl[\hbox{eq.}(3.53)\bigr]}}  \eqno(3.58)$$

The gauge choice does not affect the  two residual differential equations,
but putting $P^o=1 $ will cause some minor simplification in the form of the
differential operators.

So we have succeeded in reproducing the results in the previous paper [1]  by a
method which  was clearly influenced by the choice of the fourth coordinate $u
$
in that paper.

\

{\bf 3.4 Summary of this section.}

In Sections 3.1 and 3.2 we have succeeded in reducing the problem to
essentially a pair of coupled
ordinary differential equations for $\ell$  by a method which was
 coordinate invariant. It may be argued that we have really introduced
coordinates (the coordinate candidates) in all but name, so perhaps we should
emphasise that our method is coordinate invariant in the sense that there is no
background coordinate metric imposed, and that we do not have to adopt the
coordinate
candidates as our final coordinates. We emphasise again that  it is
structurally
imperative that we apply the commutators  to  (the equivalent of)  four
functionally independent real
 scalars  {\it to ensure that   we have a
complete system of equations.} Therefore, our prime concern was simply to
establish a complete system and we have not  been thinking of
the suitability of these quantities in their optional additional role as
coordinates;  although of course we are always
concerned with getting    the complete system of equations in a reasonably
concise and manageable form. Hence whether our coordinate candidates are the
coordinates
which give separation, decoupling and reduced order is of secondary importance,
at this stage; it is of course preferable if it happens, as it makes subsequent
work shorter.

Having
obtained explicitly a complete system we then decided to tentatively adopt the
coordinate candidates as our coordinates and to explore their
usefulness. We have seen in  Section 3.3 that this particular coordinate
choice does give separation, and this is a bonus for us; of course, we
 also want the equations to  decouple and reduce to as low an order as
possible. Although we could continue to work in the coordinates chosen in the
last section
we have pointed
out before that we have the freedom to choose  coordinates other
than the coordinate candidates;
therefore,
 we will now investigate these possibilities. From now on, having learned
something
of the structure of the residual differential equations still to be solved,
motivation for preferring a particular choice will be {\it its usefulness
not just in separation, but also in
enabling us to decouple the residual equations and to reduce the order of the
equations.}

\

\medskip

\beginsection 4. The master equations, and alternative coordinates to
coordinate candidates.

Although allowing the coordinate candidates to become the coordinates, as
 in the last section,  is the most obvious choice, it is not the only one.
 Of course, in practice, we
want to choose as coordinates those four quantities on which each of the four
derivative operators yield simple expressions, since these expressions give us
the explicit form of the operators and tetrad components in that coordinate
system; as well we recognise that  we will be left at the end
of the operator-integration procedure with a residual set of differential
equations still to be solved, and further progress towards their  solution
will depend on a suitable choice of coordinates.
However,  the particular choice made in the last section enabled us to obtain
separation of the coordinates and reduce the problem to ordinary differential
equations in the fourth coordinate $u$.  Clearly we would wish any other
coordinate choice to give us the same separation  properties; so we will
retain the first three coordinate candidates as coordinates, but postpone the
explicit choice of the fourth coordinate.
 To make a skillful and informed choice of the fourth coordinate we need more
understanding of the precise structure of the residual differential equations.
In this section we will first get a better picture of this structure,
which will motivate the eventual  explicit choice for the fourth
coordinate; so we will now  continue on directly from the end of
subsection 3.2 --- where we had reduced the problem to solving the residual
pair
of Ricci equations (3.29), subject to the inequalities (3.31) --- and
develop further the
equations (3.29) in the GHP operator notation.

\

{\bf 4.1. The master equations and redundancy.}

We have noted in Section 3.2 the crucial result that $\ell $ and $u $ are
functionally dependent. Further, when we compare the differential equations
(3.45) for $\ell  $ with those for $u $, (3.37b,3.46) we note that
$$\eqalign{\nabla_i  \ell= {\cal D}(\ell)\nabla_i u }\eqno(4.1)$$
where $$
{\cal D}(\ell)=(\bar \zeta +\zeta )^{1+m}{\tilde \Ph'\ell \over {|P^o|}}
\eqno(4.2)
$$
so  $\ell $ and  ${\cal D}(\ell)$ are functionally dependent on
$u$ and hence on each other. The operator ${\cal D}$
$$
{\cal D}={(\bar \zeta +\zeta )^{1+m} \over {|P^o|}}\tilde \Ph'
\eqno(4.3)
$$
 is simply a zero-weighted
operator  formed by scaling the
(1,1)-weighted operator $\tilde \Ph' $; this new operator  also has the
property that when it operates on a zero-weighted function of $u $, it yields
another zero-weighted function of $u $.

We now wish to write out explicitly  the two residual Ricci
equations (3.29) in this notation, and  so we first rewrite (3.40) and (3.47)
as
$$ {\tilde \Ed P^o \over {P^o}^2}= 2({\cal D}\ell+m+1)(\zeta +\bar \zeta
)^{-1}\eqno(4.4)$$
 $$
\eqalign{{2\Sigma^o\over |P^o|} & =i(\ell{\cal D}\bar \ell- \bar \ell{\cal
D}\ell
+ \ell-\bar \ell)(\bar \zeta +\zeta )^{m-1}
}
\eqno(4.5)$$
\smallskip
 The two
residual Ricci and Bianchi equations now become $$ \bar \ell \ell
{\cal D}^{3}\ell+3\bar \ell {\cal D}\ell {\cal D}^{2}\ell+\bar \ell(2m+1)
{\cal D}^{2}\ell-2 \ell {\cal D}^{2}\ell-2  ({\cal D}\ell){}^2 -(4m+2)
{\cal D}\ell-2m(m+1)=0\eqno(4.6)$$
 $$\eqalign{  &\Big\{\ell{\bar \ell}^2  {\cal D}^3 \ell+  \bar \ell
{\cal D}^2\ell\big((2m+1)\bar \ell+(2m-3)\ell+3\bar \ell{\cal D}\ell+\ell{\cal
D}\bar \ell\big) + \bar \ell({\cal D}\ell){}^2 {\cal D}\bar \ell
+(2m-3)\bar \ell  ({\cal D}\ell){}^2 \cr & \qquad\qquad
+ (2m+1)\bar \ell  {\cal D}\ell   {\cal D}{\bar \ell}
+ (4m-2)\bar \ell  {\cal D}{\bar \ell}  -(4m^2-4m-3) \ell
{\cal D}{\bar \ell} -(4m^2-4m-2) \ell \Big\}
 \cr & -  \Big\{\hbox{c.c.}\Big\} \qquad\qquad  = \quad 0
 }\eqno(4.7)$$
and the inequalities become
$$\eqalign{\bar \ell {\cal D} \ell-  \ell{\cal D}\bar \ell +
\bar \ell- \ell\ne 0
}\eqno(4.8)$$
$$\eqalign{\ell { {\cal D}^3\ell}+3{\cal D}^2\ell  {\cal D}\ell +(2m+1)
{\cal D}^2\ell \ne 0 }\eqno(4.9)$$

 These two equations (4.6) and (4.7) are not
independent, which can be seen as follows. The terms ${\cal D}^3\ell,{\cal
D}^3\bar \ell  $ can be eliminated from the second equation using the first
giving  $$\eqalign{ &\ \Big\{\bar \ell \ell\big(  {\cal D}\bar \ell   {\cal
D}^2\ell+(2m-1) {\cal D}^2\ell\big)+\bar \ell{({\cal D}\ell})^2{\cal D}{\bar
\ell} + (2m-1)\bar \ell( {\cal D}\ell)^2\cr & +(2m+1)\bar \ell {\cal
D}\ell{\cal
D}{\bar \ell}+ (4m^2-1)\bar \ell{ {\cal D}\ell}-(4m-2) \ell{ {\cal
D}\ell}-(6m^2-2m-2) \ell \Big\} \cr & - \Big\{\hbox{c.c}\Big\}  = 0
}\eqno(4.10)$$
     and it is then found that a derivative of this third equation is related
to the first by
$$\eqalign{{\cal D}\big[(4.10)\big] = (2m-1+ {\cal D}\bar \ell)
{\big[(4.6)\big]}-  (2m-1+ {\cal D} \ell)  \overline {\big[(4.6)\big]}
}\eqno(4.11)$$
In the same way as above,
 inequality (4.9) can be
simplified to  $$\eqalign{
\ell {\cal D}^2 \ell+({{\cal D}\ell})^2 + (2m+1) {\cal D}\ell+m(m+1)\ne
0}\eqno(4.12) $$
The  equations (4.6,10) and inequalities (4.8,12) for $\ell$  carry essentially
the same information as the equations (111,112)  for
$\ell$ given in [1], but the correspondence is not so obvious; for
instance we  note  the additional
explicit unknown function
$f$ given in [1], and point out that in this paper there is also
an additional
 unknown function, namely
$W$, which is implicitly built into the operator ${\cal D}$.

There are different possible ways to exploit the redundancy noted above.
 From one point of view, it means that we have essentially only two independent
real equations, e.g. (4.10) and the real part
(4.6); from another point of view we can think of the complex equation (4.6)
as the main equation with (4.10) as a supplementary equation, which is
essentially a special first integral of (a part of) (4.6). We could, at this
stage, simply write ${\cal D}=W{d\over dv}$ with $v$ as coordinate, and attempt
to
decouple and simplify the two equations, in coordinate form, making use of the
redundancy and coordinate freedom; we shall show in a subsequent paper how
this can be done, continuing on from the equations (111,112)  for
$\ell$ given in [1]. But in this paper we shall prefer to
 exploit the redundancy and simplify the
equations (4.6,10)  further,  yet remain within
the operator formalism; we shall then show that the final
coordinate choice
and decoupling follows in a very natural mannner.

We shall consider the pair of equations (4.6,10) as our `master equations',
and will simplify them in different ways in the remainder of this paper, as
well as pointing out how other approaches from these equations will enable us
to retrieve the results of earlier workers.  The equations and inequalities now
have no  terms which are functions of the three coordinates $R, \zeta, \bar
\zeta $, being functions only of the coordinate candidate $u. $ This of course
means that when we introduce explicitly our fourth coordinate we will be
dealing with a pair of ordinary differential equations in one real variable.

\medskip

{\bf  4.2. The master equations rewritten as a complex chain of first
order equations.}

We will now rewrite (4.6) and (4.10) in a  more concise and manageable form.
Defining
$$\eqalign{\lambda = & \ell ({\cal D}\ell +2m-1) \cr
\Lambda = & {\cal D}\lambda +{ 2\lambda  \over \ell   }+(m-1)(m-2)
  }\eqno(4.13)$$
we find that the  differential equation (4.6) can be written in the
   simple form
$$\eqalign{{\cal D}\Lambda =2\Lambda /\bar \ell }
\eqno(4.14)$$
while the differential equation (4.10) yields simply
$$\eqalign{
  \lambda \bar \Lambda  -(m-1)(m-2)(2  \ell+\lambda ) =  \bar \lambda
\Lambda  -(m-1)(m-2)(2  \bar \ell+\bar \lambda ) }\eqno(4.15)$$
(We can divide by $\ell $ since a non-zero $\ell $ is guaranteed by the
inequalities.)

The inequality (4.8) remains in the same form, while the inequality (4.9)
becomes very simply
$$\eqalign{
{\cal D}\Lambda \ne 0}\eqno(4.16)$$
So an equivalent  presentation of
these equations is,
$$\eqalign{{\cal D}\Lambda
= & 2\Lambda / \bar \ell
\cr
{\cal
D}\ell   = & 1-2m +  \Bigl( B +2(m-1)(m-2) \ell  \Bigr)/\ell\bigl(\bar \Lambda
   -(m-1)(m-2)\bigr)    \cr
{\cal D} B   = & \Lambda \bar \Lambda  -(m-1)(m-2)\bigl(\Lambda +\bar
\Lambda -m(m+1)\bigr)
  }\eqno(4.17)$$
where $B \bigl(= \lambda \bar \Lambda -(m-1)(m-2)(\lambda +2\ell)\bigr)$ is a
real function.

(The denominator in the second equation is non-zero, since $\Lambda$ cannot
be constant, because of (4.16); nor can $\ell$ be zero, as we have noted
above.)

At first sight this  set of first order (operator) differential
equations --- under the obvious substitutions of $\ell$ from the first into the
second,  and $B$ from the resulting equation into
the third --- appears to reduce      to a real differential operator equation
of
third order for the
   complex function $\Lambda$; however, a complication is that
since $B $ on the right hand side of the second equation is real,
this equation also carries the implicit information
$$\eqalign{ \ell\bar \Lambda
 ( {\cal
D}\ell-1+2m) & -(m-1)(m-2)( {\cal
D}\ell+1+2m)
\cr & = \bar \ell \Lambda
 ( {\cal
D}\bar \ell-1+2m)  -(m-1)(m-2)( {\cal
D}\bar \ell+1+2m)
}\eqno(4.18)$$
which complicates the deceptively simple structure  in (4.17).
Of course we also  still need to
introduce the fourth coordinate explicitly through the operator ${\cal D} $,
which will
introduce another function;   ultimately we will need to decouple the real and
imaginary parts of $\Lambda $.

We note that we have exploited the redundancy (4.11) to, in effect, replace the
7 real equations (4.13,14,15) in the 6 real unknowns $\ell,\lambda,\Lambda$
by the 5 real equations (4.17) in the 5 real unknowns $\ell,B,\Lambda$;
in particular,
it is emphasised that the introduction of the real function $B$ means that the
set of
five equations (4.17) has no redundancy.

\

   {\bf 4.3. Introducing different coordinates to achieve decoupling.}
\medskip

The above set of equations reveal a structure built around the complex function
$\Lambda  $ which we know depends  on only one variable. In
Section 3 we chose the coordinate candidate $u $ as coordinate,   and we could
of course do the same for the alternative set of equations (4.17). In that
coordinate the operator ${\cal D}= { d \over du } $, and substitution into
(4.17) will lead to two real differential equations --- of order three and two
respectively --- for the complex function $\Lambda  $; although these are of
a simpler structure than
their counterparts  obtained in Section 3, the difficulty of obtaining
decoupling --- after the separation of $\Lambda  $ into real and imaginary
parts
--- still remains.

However, we shall  illustrate two approaches
respectively in the next two sections, whereby we can obtain decoupling in a
natural way.

With our additional freedom of choice for the fourth coordinate,
 there is the
 possibility --- suggested both by the structure of the equations (4.6,10) and
the absence of $u $ explicitly in these  equations --- to choose
 the fourth coordinate in terms of $\Lambda  $. It would of course be
very attractive if we could choose $\Lambda$ itself as the coordinate;
unfortunately $\Lambda  $ is a {\it complex}\ function of the {\it real}
variable $u $ and so
$\Lambda  $
 itself cannot play the role of the fourth coordinate; however, some real
combination formed from $\Lambda $ (e.g. $\Re (\Lambda), \arg(\Lambda)$, ...)
would be  a possible
choice for the fourth coordinate. With such a choice the decoupling problem
for complex $\Lambda$ disappears, since one part of $\Lambda$ is now the
coordinate, and the other part is the real dependent unknown; there is also a
reduction in order. (It might appear that we would also be left with a
supplementary first order equation to solve for
$u$ as a function of
$\Lambda $, at the end of the analysis. In fact, as  noted above,  $u $
actually
does
 not occur explicitly in any of the  spin coefficients or Killing vector
components and so it would not even be necessary to calculate $u$ explicitly.)

 On the other hand,
if we could rearrange our equations (4.17) into a closed chain of {\it real}
equations in {\it real} functions then we could choose as our fourth coordinate
the real function
 at the top of the chain.

\

{\bf 5.  A class of real third order differential equations for all values of
$m $.}

\medskip

Let us label  our fourth coordinate, $v $, so that we can  write
$${\cal D}=W{ d \over d v }
\eqno(5.1) $$
where $W(= {dv  \over du }) $ is a function of $v $. The second of the three
equations in (4.17), with the first substituted,  becomes $$\eqalign{ 4\bar
\Lambda ^2\Bigl(\bar\Lambda- & (m-1)(m-2)\Bigr)\Bigl( \ddot {\bar \Lambda }
+(\ln W)_{,v}\dot{\bar  \Lambda}\Bigr)
\cr & + {\dot {\bar \Lambda} }^2\Bigl(WB{\dot {\bar \Lambda} }
+2(m-1)(m-2)(2m+3)\bar
\Lambda-2(1+2m){\bar \Lambda }^2\Bigr) =0
         }\eqno(5.2)$$
while the third can be written as
$$ (W B)_{,v} - WB(\ln W)_{,v}    =  \Lambda \bar \Lambda
-(m-1)(m-2)\bigl(\bar \Lambda +\Lambda -m(m+1)\bigr) \eqno(5.3)$$
where $\ \dot {} \  $ denotes differentiation with respect to $v $.

 From (5.2) we obtain
   $$\eqalign { (\ln
W)_{,v} & = \dot {\bar \Lambda}^3  \beta+{\dot
{\Lambda^{}}}^3 \bar \beta  \cr  WB & = -4\bigl (\bar\Lambda-
(m-1)(m-2)\bigr)\bar
\Lambda ^2\D{\bar \Lambda }
\beta  -4
\bigl(\Lambda-  (m-1)(m-2)\bigr)\Lambda ^2 \D\Lambda \bar
\beta}
\eqno(5.4)
$$  where
$$\eqalign{ \beta =
\Bigl(2
\Lambda ^3\ddot {
\Lambda }- & (m-1)(m-2)\bigl(2\Lambda^2\ddot {
\Lambda }-(2m+3) \Lambda
\dot{{ \Lambda }^{}}^2\bigr)-(1+2m)
  \Lambda{}^2
\dot{{ \Lambda }^{}}^2\Bigr)
\cr &
\Big/
2  \dot{ \Lambda } \dot{\bar \Lambda }\Bigl(\bigl(\bar\Lambda-
(m-1)(m-2)\bigr)\bar \Lambda ^2\dot{
\Lambda^{}} ^2 -
\bigl(\Lambda-  (m-1)(m-2)\bigr) \Lambda ^2\dot {{\bar \Lambda}^{}}
^2\Bigr)}
\eqno(5.5)$$
(We note that we can assume the denominator of $\beta $ is non-zero;
otherwise we obtain a flat space solution --- with inequality (4.16)
violated --- or the  Hauser solution [19,20,21].)

Substituting the two equations
(5.4) into (5.3) gives a real third order equation for the complex function
$\Lambda  $ of the real coordinate $v $,
   $$\eqalign{
\Bigl(\Lambda- & (m-1)(m-2)\Bigr) \Bigl(4\Lambda ^2\D \Lambda  (\D
{\bar \beta}-
 \D { \Lambda}{}^3 \bar \beta{}^2-\D { \bar \Lambda}{}^3\beta\bar \beta)
+2(2m+7) \Lambda \D { \Lambda}{}^2\bar \beta\Bigr)
\cr &
+
\Bigl(\bar\Lambda-  (m-1)(m-2)\Bigr)
\Bigl(4\bar \Lambda{} ^2\D {\bar \Lambda} ( \D
{ \beta}-
   \D {\bar \Lambda}{}^3 \beta^2-\D { \Lambda}{}^3\beta\bar \beta)
+2(2m+7)\bar \Lambda \D {\bar \Lambda}{}^2\beta\Bigr)
  \cr &
\qquad  = -\Lambda \bar \Lambda  +(m-1)(m-2)\bigl(\bar \Lambda
+\Lambda -m(m+1)\bigr) }\eqno(5.6)
$$
But we have still the freedom to choose $v $; and providing we
choose $v $ as a real function of $\Lambda  $, and make an appropriate choice
for
the dependent variable --- essentially choose it as a second, independent, real
function of $\ell$ --- then we can obtain  the resulting equation
which is of
 third order.

 As an example, we put
   $$\eqalign{  \Lambda  =  v+iX }
\eqno(5.7)$$
and write out the equation for the special case $m=1 $.    When we
substitute (5.7) into  (5.5), for this case,  we
obtain
$$\eqalign{  \beta
 = (v+iX)^2 & \Bigl(2i(v+iX)\DD X -3 (1+i\D X)^2\Bigr)
\cr & \Bigl/2(1+\D
X^2) \Bigl((v-iX)^3(1+i\D X)^2-(v+iX)^3(1-i\D X)^2 \Bigr)}
\eqno(5.8)$$
where \ $\D {}$\  denotes differentiation with respect to $v$.

When this is in
turn substituted into (5.6), which for
$m=1
$ simplifies to
$$\eqalign{
(v+ & iX)^2 (1+i\D X)\Bigl(2(v+iX)  \bigl(\D
{\bar \beta}-
 (1+i\D X){}^3 \bar \beta{}^2-(1-i\D X){}^3\beta\bar \beta\bigr)
+9  (1+i\D X)\bar \beta\Bigr) \cr &
+
(v-iX)^2 (1-i\D X)
\Bigl(2(v-iX)  \bigl( \D
{ \beta}-
   (1-i\D X)^3 \beta^2-(1+i\D X){}^3\beta\bar \beta\bigr)
+9 (1-i\D X)\beta\Bigr)
  \cr &
\qquad  = -(v^2+X^2) / 2 }\eqno(5.9)
$$
Clearly there is
no decoupling problem, and we  have  a real third order equation for the
real function $X $ of the real coordinate $v $. Picking out the  third
order expression, we can write it explicitly as
$$\eqalign{(\D\Lambda \dddot
{\bar \Lambda} -\D{\bar \Lambda} \dddot \Lambda)   = -2i\dddot X }\eqno(5.10
)$$
 We emphasise that
once we have solved this single equation (5.9) --- subject of course to the
inequalities (4.8,16) ---
       then the problem is essentially solved; once $X $ is
   obtained we can write down $\Lambda  $ from (5.7),   obtain  $(\ln
W)_{,v} $ from (5.4), and  $W $ by integration;   hence we can
obtain $\ell $ from the first of (4.17) by substituting for $W $ and
$\Lambda. $

\medskip

So, we can then write down our table for the fourth coordinate $v $,
$$\eqalign{\Ph v = & 0 \cr \tilde  \Ph' v = & { dv \over du }\tilde \Ph'u =
W|P^o|(\bar \zeta +\zeta )^{-1-m} \cr \tilde \Ed v = & { dv \over du }\tilde
\Ed
u = WP^o \ell(\bar \zeta +\zeta )^{-1}
}\eqno(5.12)$$
which gives us, in the coordinate system $R,\zeta, \bar \zeta, v $,  the GHP
operators {\it when acting on zero-weighted quantities } as,
$$\eqalign{
\Ph \equiv \ &\  \Ph (R)\,\partial_R +\Ph(\zeta)\, \partial_\zeta +\Ph(\bar
\zeta)\, \partial_{\bar  \zeta}+\Ph (v)\, \partial_v
= 2|P^o|^{-1} \partial_R
\cr
 \tilde \Ph'= &  ( \rho'^o+\bar \rho'^o)|P^o|^{-1} \partial_R+W|P^o|(\bar
\zeta +\zeta )^{-1-m}\partial _v
\cr
 \tilde \Ed= & \Bigl({2i \over |P^o|}\bigl( P^o
\Sigma^o_{,\zeta}+WP^o\ell(\zeta +\bar \zeta
)^{-1}\Sigma^o_{,v}\bigr)
 -R{P^o } \big(W \dot \ell +m+1\big) (\zeta +\bar \zeta
)^{-1}\Bigr)\partial_R \cr & \qquad\qquad+P^o\partial _\zeta
+WP^o\ell (\bar \zeta +\zeta )^{-1}\partial _v  \cr    \tilde \Ed'=  &
\Bigl({-2i \over |P^o|}\bigl( \bar P^o
\Sigma^o_{,\bar \zeta}+W\bar P^o\bar \ell(\zeta +\bar \zeta
)^{-1}\Sigma^o_{,v}\bigr)  - R{\bar P^o }\big(W\dot{\bar \ell}+m+1\big)(\zeta
+\bar \zeta )^{-1}\Bigr)
\partial_R\cr & \qquad\qquad+\bar P^o\partial _{\bar \zeta } +W\bar P^o\bar
\ell (\bar \zeta
+\zeta )^{-1}\partial _v  }\eqno(5.13)$$ where $\Sigma^o $ is given by (4.5)
and
$\ell, W $ are obtained as described above.

The Killing operators remain as in (3.50,51).

\

Finally, we emphasise  that the reduction to a real third
order equation of a real function is not only for the particular coordinate
choice $v=\Re(\Lambda ) $; we could have chosen as coordinate {\it any} real
function of $\Lambda  $ (e.g. $, \Im (\Lambda ), \arg(\Lambda ), |\Lambda|, ...
)$ and with an appropriate choice of dependent variable, achieved an equation
of the
third order. So there is a whole class of real third order equations of a real
unknown
which can easily be found from (5.6). As noted above, the leading term in (5.6)
has
the form
$(\D\Lambda
\dddot {\bar \Lambda} -\D{\bar \Lambda} \dddot \Lambda) $, and this raises the
question
whether  one of these alternative coordinate choices would enable third order
terms to cancel, and yield an equation of second order. For the special
coordinate choices just mentioned it is easy to see that such reduction does
not occur, and indeed we shall show in a subsequent paper that there is no
possible way to choose our coordinate as a real function of $\ell$ in order to
achieve reduction of order. Of course this does not mean that there may not be
some other choices of coordinate for which reduction of order can occur.

\

\medskip

{\bf 6. A concise third order equation for the case $m=1. $}

\medskip

In this subsection we reduce the system of complex equations (4.17) to a real
system, which form a closed chain of real
first order
(operator) differential equations.  So as not to obscure the technique by
details
we consider the special case  $m=1 $. Beginning with the one real equation from
(4.17), $$\eqalign{
{\cal D}B & =C}\eqno(6.1)$$
where
$$\eqalign{B & = \lambda \bar \Lambda \cr C & = \Lambda \bar \Lambda
}\eqno(6.2)$$
we use the other equations in (4.16) to obtain the two real equations
$$\eqalign{
{\cal D}C & =2CE   }\eqno(6.3)$$
and
$$\eqalign{
{\cal D}E &  = -{ B  \over \bar \Lambda \ell^3 }-{ B  \over \Lambda \bar \ell^3
}
+{ 1 \over \ell^2 }+{ 1 \over \bar \ell^2 }
}\eqno(6.4)$$where
$$\eqalign{E & = ({ 1 \over \ell } +{ 1 \over \bar
\ell })
}\eqno(6.5)$$
with $B,C,E\ne 0 $.

We could obtain the next equation in the chain by  operating with ${\cal D} $
on the right hand side of (6.4); but the calculations are shorter if
 we introduce the complex function $F$ by
$$\eqalign{ F= \ell^2\bar \Lambda }\eqno(6.6)$$
and note that not only can we write out the right hand side of (6.4) in terms
      of real $B,C,E $ and complex $F$, but we also have, from
(4.17), the very simple and useful result    $$\eqalign{{\cal D}F &
=2B}
\eqno(6.7)$$
Since $B $ is real we can put
$$\eqalign{ A+ik=F=\ell ^2\bar \Lambda
}\eqno(6.8)$$
where $A $ is a real function, but $k$ is a constant. So we have now obtained
the
following  closed chain of real first order equations of real functions:
$$\eqalign{ {\cal D}A & = 2B \cr
{\cal D}B & =C\cr
   {\cal D}C & =2CE\cr
  {\cal D}E &  = -{ B \over A^2+k^2 }\Bigl( A E \pm k \sqrt{E^2-4\sqrt{{ C
\over
A^2+k^2}}}\Bigr) +E^2-2 \sqrt{{ C \over
A^2+k^2}}
} \eqno(6.9)$$
Although we now have four real differential equations compared to
the (equivalent of) five
real differential equations in (4.17) we have not lost any information; the
 missing differential equation is simply  ${\cal D} k =0$; and by using only
real functions all the information is explicit within the chain.
Substituting the first three equations of (6.9) into the fourth one
will clearly give a fourth order operator equation; however, an
appropriate choice of coordinate will reduce the corresponding
coordinate equation to third order.  Let us  choose $ A
$ as our fourth coordinate so that from the first equation the operator  ${\cal
D}  $ is given by
 $$\eqalign{ {\cal D}=2B { d \over d A } }\eqno(6.10)
$$ and
substituting  from the first three equations of (6.9) in the fourth equation
gives a real third order differential equation for $B $, a real function of $A
$, which can be presented in the comparatively concise form,
 $$\eqalign{
2{{S}}{\dot{S}} & \dddot {S}+{\dot{S}}^2\ddot{S}-3{{S}}{\ddot{S}}^2 \cr & =
-{{\dot{S}}\over (A^2+k^2)}\Bigl(A{{S}}{\ddot{S}}
\pm k   \sqrt{{S}^2{\ddot{S}}^2-  4  {S}{\dot{S}}^2\sqrt{{
\dot{S} \over A^2+k^2}} }\ \
 \Bigr)  -2 {\dot{S}}^2 \sqrt{{\dot{S}}\over {A^2+k^2}}}
\eqno(6.11)
$$
where $\ \dot{} \  $ denotes differentiation with respect to $A $ and
$$\eqalign{
{S}= B^2}\eqno(6.12)$$
The constant parameter $k $ can be absorbed by the following relabelling
$$\eqalign{\tilde A = & A/k \cr
\tilde S = & S/k
}\eqno(6.13)$$
giving
$$\eqalign{
2{{\tilde S}}{\dot{\tilde S}}\dddot {\tilde S}+{\dot{\tilde S}}^2\ddot{\tilde
S}-3{{\tilde S}}{\ddot{\tilde S}}^2= -{{\dot{\tilde S}}\over ({\tilde
A}^2+1)}\Bigl({\tilde A}{{\tilde S}}{\ddot{\tilde S}} +  &
\sqrt{{\tilde S}^2{\ddot{\tilde S}}^2-  4  {\tilde S}{\dot{\tilde S}}^2
\sqrt{{ \dot{\tilde S} \over {\tilde A}^2+1}} }\ \
 \Bigr)  -2 {\dot{\tilde S}}^2 \sqrt{{\dot{\tilde S}}\over
{{\tilde A}^2+1}}}\eqno(6.14)$$ where $\ \dot{} \  $ now denotes
differentiation
with respect to the new coordinate
${\tilde A} $.

It is clear that once ${\tilde S}$ is obtained the problem is essentially
solved:
   $C,E $ can be obtained from (6.9) by differentiation, and can be combined
algebraically to give  $\ell $,
$$\eqalign{
{2\sqrt {k}\over \ell}= { \sqrt{{\tilde S}} \ddot {\tilde S}\over
{\dot{\tilde S}}^2   }+ \sqrt{ {{\tilde S}{\ddot{\tilde S}}^2\over {\dot{\tilde
S}}^2}-
4  \sqrt{ { \dot{\tilde S} \over {\tilde A}^2+1}
}
}  }\eqno(6.15)$$
The table of GHP operators in this coordinate system,
$R,\zeta ,\bar \zeta ,{\tilde A} $ is obtained from (5.10) where $v $
is replaced by $A $ and then by ${\tilde A}$,  and $W $ is replaced by $B $,
and then by
${\tilde S}$.

We emphasise that the inequalities (4.8,16) must also be satisfied;
this constraint prevents us from making the simple choice $k=0 $,
because it violates the first of these inequalities.

\smallskip

We could have chosen other coordinates e.g. $B $ or $C $ or combinations,
and in the same manner obtained alternative third order equations.

Finally, we note  the relative simplicity of the version of the final equation
obtained here as compared to the version, for $m=1, $ in the last section.

\

\medskip

{\bf 7. Summary.}

The original purpose of this paper was not primarily to try and make
significant
new progress in the NT problem; it was rather to set out in detail new insights
into the GHP operator-integration approach --- whose principles are applicable
in
very general contexts --- and  then to illustrate these insights by applying
them
to the  NT problem.   But, as well as the NT problem
providing an ideal laboratory for the demonstration of our GHP
operator-integration approach, we have been able to obtain  new
insights into the NT problem itself --- extending existing results and
suggesting new approaches.

The major new insight regarding the GHP operator-integration approach is the
important fact that, although spacetimes admitting Killing vectors do not
immediately supply --- from the spin coefficients and Riemann tensor components
--- the four functionally independent scalars which integration within the GHP
formalism  demands,  the missing scalars may be supplied in a very simple and
natural way --- by also using the (tetrad components of the) Killing vectors.
The
relationship in the GHP formalism between Killing vectors and functionally
independent scalars, as  well as that between  homothetic Killing vectors
  and separability, are  very important topics which we have only touched on
here; deeper implications for the GHP formalism will be developed in a
separate paper.

An additional insight regarding the GHP operator-integration approach is
that, even after we have reduced the problem to the residual
ordinary differential operator
equations, there are advantages to remaining within an operator formalism
rather than immediately using coordinates explicitly. We have shown that, by
constructing a closed
 chain of real first order ordinary differential operator equations,
we can avoid the difficulties associated with decoupling, and can make the
final
coordinate choice in an efficient manner.

\medskip

We chose the $G_2 $ case of the NT problem as our
illustration because of its non-trivial nature and because of the wide range of
mathematical procedures  necessary for its  simplification.
 The intention was that  the advantages and strengths of our method
be thoroughly illustrated and tested. Much work has already gone into this
particular problem, and  many
insights have already been gained --- although often in a rather narrow manner
specific to a particular formalism and coordinate system. In this paper, we
have
been able to understand the procedures of separability, redundancy, decoupling
and reduction of order in a very general manner, in the context of the GHP
formalism; we have also seen how it is possible to retain our coordinate choice
to the very last step when it can be used to  simplify the final equations
in the most advantageous way.
This we believe is the strength of this GHP operator approach; separability,
redundancy  even decoupling can be exploited, still retaining some coordinate
freedom, and then at the very last stage when everything has been reduced to
the
decoupled residual ordinary differential operator equations this freedom can be
exploited in such a manner
 as to present the final equations in the most reduced
and/or manageable form.

Other workers, [15,16,17,18] who have reduced this problem to third
order real differential equations have used very different formalisms, and
special  techniques, and
 there is no simple relationship between them. However, it is possible to
obtain these very different equations from our master equations:

\smallskip
(i) The
form of the residual real third order ordinary differential equation for
$h(\varphi) $ obtained by Ludwig and Yu [17] follows in a very concise
manner from our master equations (4.6,10), by making the special choice of
coordinate
$\varphi = \arg (\ell) $ and of independent variable $h= {d \varphi\over
du}|\ell| $. We note,  that unlike in the approach in Section 5, the particular
coordinate choice is crucial;  most other coordinate choices built on $\ell$
give fourth order equations.

\smallskip
   (ii)  For arbitrary values of the parameter $m $, Herlt [16] has first of
all
 reduced the problem essentially to a real third order ordinary differential
equation for a complex function $g $, which he then transforms to a first
order  integro-differential equation for the real function $\varphi(u) $ by
defining the coordinate $u $ by $$\eqalign{g=g_{,u}e^{i\varphi(u)}
}\eqno(7.1)$$
Finally,  he is able to obtain a real third order ordinary differential
equation for a real function  $\chi(\varphi) $ defined by
$$\eqalign{ \chi(\varphi)= \int \sin \varphi\ du - 2\varphi(u)  }
\eqno(7.2)$$
It is
easy to see that Herlt's function $g $ corresponds essentially to our function
$\lambda  $, and so we can retrieve his result by first rearranging our master
equations (4.6,10), and then following his procedure. However, it is
also possible to obtain other third order equations  by constructing different
coordinate choices around $\lambda$ (equivalently Herlt's $g$) than the
choice (7.1).

For the special case $m=1 $ (corresponding to N=2, in the notation of [16])
Herlt has obtained a particularly concise form for the residual real third
order differential equation, just as we have also, for this special case ---
in
Section 6. Although there
are obvious similarities in structure, some manipulation still needs to be
carried out to show the direct equivalence of the two equations.

 \smallskip
(iii) Chinea [15] --- using an elegant approach with matrix
valued differential forms --- has reduced the NT problem with two commuting
Killing vectors (i.e. $m=0$) to a single complex second order differential
equation  for a complex function. If we  choose  $B $ as our coordinate,
then after substituting the first and third into the second of equations (4.17)
we obtain a
complex second order differential equation for arbitrary values of the
parameter $m $;  for the special case $m=0 $ this equation is
$$(\Lambda\bar\Lambda-2\Lambda-2\bar\Lambda)(2\bar\Lambda
\ddot{\bar\Lambda}-\dot{\bar\Lambda}{}^2)=
2\dot{\bar\Lambda}(2\dot{\bar\Lambda}\Lambda-2\bar\Lambda\dot\Lambda
+{\bar\Lambda}^2\dot\Lambda)-B(\Lambda\bar\Lambda-2\Lambda-
2\bar\Lambda)^2 \dot {\Lambda}^3
/2{\bar\Lambda}^2
\eqno(7.3)$$
where \ $\dot{}$\ means differentiation with respect to $B$.
This equation is similar in structure
to, but is not yet as neat as that of Chinea [15]. However, for non-commuting
Killing vectors  the two special cases $m=1,2 $ have the
remarkably simple forms, given respectively by
$$\eqalign{\Lambda \ddot
{\bar \Lambda} +   \dot { \Lambda}\dot {\bar \Lambda} = (2\bar
\Lambda -B\Lambda \dot {\bar \Lambda})
\Lambda \dot {\bar \Lambda}^2/4\bar \Lambda^2  }
\eqno(7.4)$$
$$\eqalign{\Lambda \ddot
{\bar \Lambda} +   \dot { \Lambda}\dot {\bar \Lambda} = (6\bar
\Lambda -B\Lambda \dot {\bar \Lambda})
\Lambda \dot {\bar \Lambda}^2/4\bar \Lambda^2  }
\eqno(7.5)$$
Chinea
was able to transform his complex second order equation, by a number of
coordinate changes, to a very complicated  real third order ordinary
differential equation for a real function of a real variable.  We could do the
same for the  equations above (and also for the equation for arbitrary $m $),
but the
   resultant equations seem more complicated than the versions obtained  in the
previous sections.

\smallskip
(iv) Finley et al. [18], beginning from the point of view of groups of point
transformations, also reduce the problem to one real third order ordinary
differential equation for a real function of a real variable $q(y) $. They also
define a function $\Psi(y) $,
which is second order in $q $ and
point out that the complicated third order equation can be rewritten as a
comparatively
simple, but interesting, first order expression for the real function $\Psi(y)
$
 --- spoiled only by
{\it one} additional  term
   explicit in $q(y) $. McIntosh [23] has also found a similar interesting
first order expression, again, unfortunately with {\it one} additional
complicated term.  As Finley et al. point out, these first order expressions
have
suggestive  symmetries, which may lead to further significant simplifications.
Essentially the idea being followed  is to write the very complicated third
order equation as a coupled pair of equations; in particular one of
these equations should be simple enough to enable solutions for it to be
found, yet complicated enough that the solutions are not trivial, and also
that the second equation in the coupled pair should be significantly
simplified.

In fact, our versions of the
master equations in both (4.17) and (6.9) are in a form which   enables us to
experiment in precisely this manner. For instance,
let us choose $A$ as coordinate and rewrite
 the last equation in the set (6.9) as
$$
2 X_{,A} - X^2 +{A X\over A^2+k^2}= Y
\eqno(7.6)$$
where we have relabelled
$
X = E/B$,
and where $Y$ is given from (6.9) by
$$
Y = {XC\over 2 B^2}
 \pm{ k \over A^2+k^2 } \sqrt{X^2-{4\over B^2}\sqrt{{ C \over
A^2+k^2}}}-{2\over B^2} \sqrt{{ C \over
A^2+k^2}} \eqno(7.7)$$
 From this equation we can easily find an expression for $X$  in
terms of $B,C,Y$ and the coordinate $A$. If we differentiate this expression
twice we get two new expressions respectively in $ Y_{,A},Y,B,C,$ and
$ Y_{,AA},Y_{,A},Y, B,C$; by eliminating $B,C$ between these three
expressions we are left with an expression for $X$  in
terms of $Y,Y_{,A},Y_{,AA}$ and the coordinate $A$ i.e. an expression for $X$
which is second order in $Y$.

The structure of the left hand side of (7.6) is very similar to the
interesting first order equations obtained respectively by Finley et al. and
McIntosh.

   \smallskip
(v) Finally we point out that the general third order equation which we
obtained
in Section 5 can also be obtained using the coordinate method of our earlier
paper, [1]. By a suitable coordinate transformation, and appropriate
combination
of equations (111) in [1], we can obtain [5.6], after a lengthy calculation.

   \smallskip

We shall show in a
further paper the full details of how all of the existing equations, together
with some simpler new ones, can
   be obtained from our master equations.

 It is clear that there are many  versions of a
final  third order
equation, and so it is certainly possible that more manageable ones exist;
there
is even the possibility that  a second order equation may still be found.  We
believe that the approach in this paper, which has the power to give us a
unifying picture of such different approaches,  supplies us with very powerful
equipment for this ongoing task.

\

{\bf APPENDIX: \   SIMPLIFICATION OF GHP EQUATIONS IN  INTRINSIC
TETRADS.}

\medskip
{\bf A.1.  Spacetimes admitting a Killing Vector.}

\smallskip
We emphasise that in [12] the  system of equations given was for the presence
of a homothetic Killing vector, while in [8,9] only the presence
of a  Killing vector is considered.

 In this subsection we are  only considering Killing vectors, so
when we refer
 to the equations in [12] we understand that we have made the substitution
$\phi=0$ which reduces the conformal Killing equations  to the
Killing equations.

In [8,9] the Killing equations consist of the seven complex and the two real
equations  (A.8a,b,c,9a,b,10a,b,13,14) while in [12]
the  Killing equations
consist of the eight complex and four real equations ((21), (21)$'$, (21)*,
(21)$'$*, (22), (22)$'$, (22)*, (22)$'$*, (23), (23)$'$,
(23)*,
(23)$'$*).

The discrepency in the number of equations is easily explained when we realise
that (A13) in [8,9]
has simply been split in [12] into the two equations (22)$'$ and (22) by the
introduction of an arbitrary real quantity ${\cal P}$; similarly (A14) in [8,9]
has simply been split in [12] into the two equations (22)* and (22)$'$* by the
introduction of an arbitrary real quantity ${\cal P^*}$. Finally in [12],
(21)$'$* is simply the complex conjugate of (21)*; in [8,9] only (A.8b), the
counterpart to (21)* in [12], is given explicitly.

But there is a second apparent discrepency.  In [12] there occurs in (23) and
(23)$'$*((23)$'$ and (23)*) an arbitrary complex quantity ${\cal Q}$ (${\cal
Q}'$); no such arbitrary quantities occur in the counterpart equations
(A9a,b) and (A10a,b) in [8,9].
However,  it is easy to see that when we apply the vacuum Type N restrictions
with the particular tetrad gauge used in [12] to the complete system of
equations
in [12], that it follows that
$$ {\cal Q}=0= {\cal Q'}
\eqno(AI.1)$$
(In fact, in [12] this calculation was carried out for the more general system
--- admitting a {\it homothetic} Killing vector --- resulting in (AI.1).

{\it Therefore, the apparent discrepency between the two systems of equations
in [12] and [8,9] respectively, in the presence of a Killing
vector is resolved --- for the particular class of spacetimes under
consideration in this paper.}

 However,  we  know from Held's argument that such simplifications can be made
in much more general circumstances. This is a very fundamental property for
spacetimes admitting  homothetic Killing vectors, and has very important
consequences for the GHP formalism.  We will deal with this topic elsewhere;
here we will just illustrate other more general cases where this simplification
can be made.

\medskip

{\bf  (i)}\ For instance, suppose we choose $l^i$ along the principal null
direction of the Weyl tensor, so that
$$\Psi_0=0
$$
Then   we can deduce that
$$
{\cal Q}=0
$$
provided that at least one of the $\Psi_n$$'$s $(n=0,. . . ,4)$ is nonzero, by
an iterative argument
using in order (43), (44), (45), (44)$'$, from [12]. (See [22] for a related
type of discussion.). Similarly we can choose $n^i$ along the principal null
direction of the Weyl tensor, so that  $$\Psi_4=0
$$
and this is compatible with
$$
{\cal Q'}=0
$$
(We note that the choices $\Psi_3=0=\Psi_1$ also lead to the same result.)

{\bf  (ii)}\ For algebraically special spaces if we choose $l^i$ such that
$$\Psi_0=0=\Psi_1$$
then, as in (i),
$$
{\cal Q}=0
$$
We can use the null rotation about $l^i$ to put
$$
{\cal Q'}=0
$$

In an analogous manner we could fix $l^i$ with respect to the Ricci tensor
etc.,
[22].

\medskip

{\bf A.2.  Spacetimes admitting a Homothetic Killing Vector.}

\smallskip

We now consider the set of equations in [12] in the presence of a homothetic
Killing vector, and as noted above, it has been shown in [12] that (AI.1)
follows when we apply the vacuum Type N restrictions
with the particular tetrad gauge used in [12]. Therefore, for this special
case,  we can combine the 12 equations in [12] into the following 9 equations,
$$\eqalign{\tilde  \Ph' a & = -\kappa' \bar c- \bar \kappa' c \cr
\bar \rho\tilde \Ed a & = -\bar \kappa'b +(\bar \rho'-\rho')\bar c \cr
\Ph b & = 0 \cr
\bar \rho\tilde \Ed b & = (\rho-\bar \rho)\bar c  \cr
\Ph c & = -\bar \rho c \cr
\tilde \Ph' c & = - \rho' c \cr
\rho\tilde \Ed' c & = 0 \cr
\tilde \Ph'b + \Ph a & = \phi \cr
\rho\tilde \Ed' \bar c+\bar \rho\tilde \Ed c &  = -\phi -(\rho'+\bar
\rho')b-(\rho+\bar \rho)a}\eqno(AI.2)$$  These equations are just the
homothetic
Killing vector generalisations of the analogous Killing vector equations
(3.1)--(3.9) in [8], specialised to vacuum Type N spacetimes.

We could have derived these equations for much more general classes of
spacetimes, using a generalisation of the geometric argument used by Held in
[8];
however,  this would involve us in deeper questions than we wish to consider
here, and the above equations are sufficient for our purposes.

\

\

{\bf Acknowledgements.}

One author (B.E.) would like to thank the Mathematics Department of the
University of Alberta for its hospitality while part of this work was being
carried out. He is also grateful for the  support of the Swedish Natural
Science Research Council for most of the period of this work.  The other author
(G.L.) wishes to thank the Department of Mathematics, Link\"oping University
for
its hospitality during part of the time this work was in progress and the
Swedish Institute for support during that visit. He is also grateful for the
continuing financial support by the Natural Sciences and Engineering Research
Council of Canada. We would also like to give special thanks to Alan Held for
challenging us to tackle this problem in the GHP formalism; we also appreciate
his stimulating suggestions and his constructive criticisms both regarding
this particular paper and other related work.

\

\

{\bf References.}

1. Ludwig, G., and Edgar,  B.   `Integration in the GHP formalism I: A
coordinate approach with
applications to twisting Type N spaces.' {\it To be published in G. R. G.}

2. Geroch, R., Held, A., and Penrose, R. (1973). {\it J. Math. Phys.,} {\bf
14,}
874.

3. Newman, E.T. and Unti, T. (1962). {\it J. Math. Phys.,} {\bf 3}, 891.

4.  Newman, E.T. and Unti, T. (1963). {\it J. Math. Phys.,} {\bf 4}, 1467.

5.  Newman, E.T. and Penrose, R. (l962). {\it J. Math. Phys.,} {\bf 3,} 566.

6. Held, A. (1974). {\it Commun. Math. Phys.,} {\bf 37,} 311.

7. Held, A. (1975). {\it Commun. Math. Phys.,} {\bf 44,} 211.

8. Held, A. (1976). {\it G. R. G.,} {\bf 7,} 177.

9. Held, A. (1976). {\it J. Math. Phys.,} {\bf 17,} 39.

10. Held, A. (1985). In {\it Galaxies, Axisymmetric Systems and
Relativity} (ed. M.A.H. MaCallum), Cambridge University Press. p.208.

11. Edgar, B. (1992). {\it G. R. G.,} {\bf 24,} 1267.

12. Kolassis, C. and Ludwig, G. (1993). {\it G. R. G.,} {\bf 25,} 625.

13. Stewart, J.M. and Walker, M. (1974). {\it Proc.
Roy.  Soc. A} {\bf 341,} 49.

14. McIntosh, C. B. G. (1985). {\it Class. Quantum Gravity,} {\bf  2}, 87.

15. Chinea, F. J. (1988). {\it Phys. Rev. D} {\bf  37}, 3080.

16. Herlt, E. (1991). {\it G. R. G.,} {\bf  23}, 477.

17. Ludwig, G. and Yu, Y. B. (1992). {\it G. R. G.,} {\bf  24}, 93.

18. Finley, J.D. III, Pleba\'nski, J. F. and Przanowski, M.
(1994). {\it Class. Quantum Gravity,} {\bf  11}, 157.

19. Hauser, I. (1974). {\it Phys. Rev. Lett.} {\bf  33,} 1112.

20. Hauser, I. (1978).  {\it J. Math. Phys.,} {\bf  19,} 661.

21. Ernst, F.J. and Hauser, I. (1978). {\it J. Math. Phys.,} {\bf  19,} 1816.

22. Kolassis, C. A. and Santos, N.O. (1987). {\it Class. Quantum Gravity,} {\bf
4}, 599.

23. McIntosh, C. B. G. (1995). Seminar at GR14 , Florence, and private
communications.

\end